\documentclass{aa}
\usepackage{graphicx}
\usepackage{txfonts}

\begin{document}

\title{{\sc SigSpec}}

\subtitle{I. Frequency- and Phase-Resolved Significance in Fourier Space}

\author{P. Reegen\inst{1}}
\institute{Institut f\"ur Astronomie, Universit\"at Wien, T\"urkenschanzstra\ss e 17, 1180 Vienna, Austria, \email{reegen@astro.univie.ac.at}}

\date{Received October 19, 2006; accepted March 6, 2007}

\abstract
{Identifying frequencies with low signal-to-noise ratios in time series of stellar photometry and spectroscopy, and measuring their amplitude ratios and peak widths accurately, are critical goals for asteroseismology. These are also challenges for time series with gaps or whose data are not sampled at a constant rate, even with modern Discrete Fourier Transform (DFT) software. Also the False-Alarm Probability introduced by Lomb and Scargle is an approximation which becomes less reliable in time series with longer data gaps.}
{A rigorous statistical treatment of how to determine the significance of a peak in a DFT, called \sc SigSpec\rm , is presented here. \sc SigSpec \rm is based on an analytical solution of the probability that a DFT peak of a given amplitude does not arise from white noise in a non-equally spaced data set.}
{The underlying Probability Density Function (PDF) of the amplitude spectrum generated by white noise can be derived explicitly if both frequency \em and \rm phase are incorporated into the solution. In this paper, I define and evaluate an unbiased statistical estimator, the ``spectral significance'', which depends on frequency, amplitude, and phase in the DFT, and which takes into account the time-domain sampling.}
{I also compare this estimator to results from other well established techniques and assess the advantages of \sc SigSpec\rm , through comparison of its analytical solutions to the results of extensive numerical calculations. According to those tests, \sc SigSpec \rm obtains as accurate frequency values as a least-squares fit of sinusoids to data, and is less susceptible to aliasing than the Lomb-Scargle Periodogram, other DFTs, and Phase Dispersion Minimisation (PDM). I demonstrate the effectiveness of \sc SigSpec \rm with a few examples of ground- and space-based photometric data, illustratring how \sc SigSpec \rm deals with the effects of noise and time-domain sampling in determining significant frequencies.}
{}

\keywords{methods: data analysis -- methods: statistical}

\maketitle

\section{Overview}\label{S Overview}

In this paper I provide a brief introduction to Fourier methods in astronomical time series analysis (Section\,\ref{S Introduction}), outline existing statistical approaches (Section\,\ref{SUB Statistical aspects}), and address the major weaknesses in available techniques (Section\,\ref{SUB Problems}). 

While Sections\,\ref{S Introduction} to \ref{SUB Problems} contain previously available (textbook) information, I introduce a new and unbiased reliability criterion ({\em spectral significance}) based on theoretical statistics in Section\,\ref{S Amplitude PDF for non-equidistant sampling}. This section also addresses the correspondence between the spectral significance and other reliability estimators. Section\,\ref{S Numerical tests} is devoted to the comparison of the analytically deduced spectral significance to the results of numerical simulations. Finally, the results of comparative tests of spectral significance computation vs.~other period detection methods are presented.

An example for the practical application of the new method vs.~a widely used standard procedure is provided in Section\,\ref{S Practical application}. It may be useful for the non-mathematically oriented reader who is mainly interested in how the technique performs in `real life'.

Further topics (Section\,\ref{S Further topics}) are the application of statistical weights, the fact that a time series consists of individual subsets, and potential problems with colored vs.~white noise.

\section{Introduction}\label{S Introduction}

In {\em Fourier Analysis} a continuous function of time over a finite time interval is expanded into a series of sine waves. These waves represent a superposition of an oscillation at a fundamental frequency and a discrete, generally infinite set of overtones. The fundamental frequency is determined by the reciprocal time interval width. All other frequencies correspond to integer multiples of this fundamental. The knowledge of the amplitudes and phase angles for all frequency components permits one to entirely recover the given function in the time domain.

Practical applications (such as astronomical observations) generally deal with discrete sets of measurements ({\em time series}) rather than continuous functions of time and, on the other hand, consider the Fourier Spectrum as a continuous function of frequency rather than a discrete dataset. This leads to the {\em Discrete Fourier Transform (DFT)}. It allows one to determine the dominant frequencies of the observed physical process with a higher frequency resolution than is possible with Fourier Analysis.

Motivated by the desire to understand physical oscillations, the scientist is interested in a couple of eigenfrequencies and the exact determination of related amplitudes and phases rather than the complete signal recovery. In practical applications, these are considered to correspond to local maxima ({\em peaks}) in the {\em amplitude spectrum}. A widely held strategy is to 
\begin{enumerate}
\item compute an amplitude spectrum for the given dataset,
\item identify the maximum amplitude within the frequency range of interest,
\item decide whether this amplitude is `significant',
\item subtract the corresponding sinusoidal signal from the time series, and
\item use the residuals after subtracting the fit from the time series for the next iteration.
\end{enumerate}
This procedure is to be understood as a loop, terminated if the maximum amplitude is not considered significant any more. The result of these consecutively performed prewhitenings is a list of frequencies, amplitudes, and phase angles, plus a residual time series (hopefully) representing the pure observational noise. In fact, part of this noise is due to measurement errors, but frequently merged with signal components the amplitudes of which are too weak to be detected. As an example, the number of photometrically resolved frequencies in the $\delta$\,Sct star 4\,CVn increased from 5 to 34 between 1990 and 1999 (Breger et al.~1990, 1999).

In many cases, the results of the prewhitening are subject to a multiperiodic least-squares fit (e.\,g., Sperl~1998; Lenz \& Breger~2005). This represents a fine tuning to adjust frequencies, amplitudes, and phases to a minimum rms residual, but may lead to the exclusion of some terms, or inclusion of new terms.

The eigenfrequencies, amplitudes, and phases of stellar oscillations provide fundamental information on the distribution of mass and temperature, the radiative and convective energy transport, or abundances of elements in the stellar atmosphere and help to determine fundamental parameters such as mass, radius, effective temperature, rotational velocity, and age of a star.

\section{Statistical aspects}\label{SUB Statistical aspects}

In practical applications, a signal is not only of stellar origin but a superposition of the information received from the star, instrumental (pseudo-)periodicities (e.g., invoked by thermal effects), and transparency variations in the Earth's atmosphere. To eliminate the third component, obser­vations with instruments aboard of spacecraft have been used increasingly by the as­tronomical community during the past two decades. Unfortunately, stray light scattered from the illuminated surface of the Earth introduces additional quasi-periodic artifacts that are not easy to handle and hence represent the major constraint to the accuracy of space-based data acquisition (Reegen et al.~2006).

An unbiased criterion to decide, whether a peak amplitude is generated by noise or represents an intrinsic variation, is important, because the choice of the most significant peak determines all further iterations in the prewhitening sequence. A falsely identified signal component usually perturbs all results obtained subsequently.

In addition, Scargle (1982) points out that Gaussian noise in the time domain may produce a peak of arbitrary amplitude in the DFT spectrum. Since there is no natural upper limit to amplitudes produced by noise, the danger of misinterpretation is imminent, and the `significance' of an examined peak needs to be described by a probability distribution.

Of course, the probability that white noise produces a low-amplitude peak is higher than that for a high-amplitude peak. In other words, the {\em False-Alarm Probability} that the highest peak in an amplitude spectrum is an artifact due to noise is lower for higher amplitudes. This definition is reasonable because it relies on the highest available peak, since one cannot trust any peak, if even the one with the highest amplitude is unreliable. In this perspective, the False-Alarm Probability appears to be a good criterion of whether to believe in the presence of a signal in a given dataset.

The statistical description of the False-Alarm Probability relies on a {\em Probability Density Function (PDF)}, which is the continuous version of a histogram, as the bins of which become infinitesimally small. Thus the False-Alarm Probability is an integral over the PDF. In many statistical applications, probability distributions may easily be described by the PDF, but an analytic expression of the integral does not exist, as e.\,g. for the Gaussian distribution. Hitherto many statistical problems are solved in terms of PDFs rather than of the corresponding cumulative quantities.

The PDF of the amplitude spectrum generated by pure Gaussian noise at equidistant sampling was deduced by Schuster (1898; also Scargle 1982). For non-equidistant sampling, the {\em Lomb-Scargle Periodogram} (Lomb 1976) is claimed to provide a better statistical behavior (Scargle 1982). Koen (1990) examined the application of the Fisher test (1929) to the Lomb-Scargle Periodogram to estimate peak significance, and Schwarzenberg-Czerny (1996) combined Lomb's solution with the analysis of variance method (Schwarzenberg-Czerny 1989) to obtain a powerful period search technique for non-equidistant data.

A widespread, but purely empirical approach is the consideration of a peak in the amplitude spectrum to be `real', if its amplitude signal-to-noise ratio exceeds a given limit. Breger et al.~(1993) suggest a value of $4$. Numerical simulations for $19\, 300$ synthetic time series by Kuschnig et al.~(1997) return an associated False-Alarm Probability of $10^{-3}$ for $1\, 000$ data points.

An alternative way of determining periods was introduced by Lafler \& Kinman (1965) and statistically examined by Stellingwerf (1978). The {\em Phase Dispersion Minimization} method is based on the assumption that the correct period would produce a phase diagram with the lowest intrinsic scatter. This algorithm is a powerful tool especially for non-sinusoidal periodicities, e.\,g., if the stellar surface structure is observed photometrically as the star rotates. In this situation, the advantage compared to Fourier techniques is the simultaneous treatment of all overtones that determine the shape of the signal. 

In the case of multiperiodic variability, PDM loses accuracy, since the phase diagram for one frequency is contaminated by all others, unless they are integer multiples of this frequency.

\section{Problems}\label{SUB Problems}

A closer examination of the fundamental properties of the DFT leads to the following issues:
\begin{enumerate}
\item The peak frequency in the amplitude spectrum of a single sinusoidal signal is not recovered correctly (Kovacs 1980). This is a side effect of the restriction to a finite time interval rather than a property of non-equidistant sampling. Also the Lomb-Scargle Periodogram suffers from systematic peak frequency displacements (see \ref{SUB Accuracy of peak frequencies}).
\item The `entire' spectrum is defined by the {\em Nyquist (Critical) Frequency} as the upper limit for equidistantly sampled data (Whittaker 1915; Nyquist 1928; Kotelnikov 1933; Shannon 1949; Press et al.~1992). In the case of non-equidistant sampling, there is no unique definition of such a limit. Horne \& Baliunas (1986) suggest the average sampling interval width for the calculation of the Nyquist Frequency; Scargle (1982) and Press et al.~(1992) promote the minimum sampling interval; a variant discussed by Eyer \& Bartholdi (1999) is to pad the entire dataset with zeros to achieve equidistant, but much denser sampling in the time domain and to take the resulting sampling interval as a Nyquist Frequency estimator. The most promising method was introduced by Sperl (1998): each sampling interval of the non-equidistant series is considered responsible for its own, individual Nyquist Frequency. In a subsequent step, a histogram of all individual Nyquist Frequencies helps to decide where to define a healthy upper frequency limit.
\item Periodic gaps in the time-domain sampling introduce periodicities in the Fourier Spectrum. In this case, a single sinusoidal signal is represented by a peak that is accompanied by several aliases. In order to overcome the systematic effects of non-equidistant sampling in the frequency domain, Ferraz-Mello (1981) examined the possibility of incorporating the sampling properties into the DFT amplitude spectrum by introducing appropriate statistical weights. His study led to the design of harmonic filters, which may help to improve the selection of peaks. A different approach is the consideration of the entire `comb' of aliases instead of the peak with the highest amplitude only (Roberts, Lehar \& Dreher 1987; Foster 1995).
\end{enumerate}

The goals of the present paper are
\begin{itemize}
\item to deduce the depedence of the Probability Density Function of non-equidistantly sampled Gaussian noise on frequency and phase (Section\,\ref{S Amplitude PDF for non-equidistant sampling}),
\item to introduce spectral significance as a measure of  False-Alarm Probability (\ref{SUB Significance}),
\item to compare the theoretical solution with the results of numerical simulations (\ref{SUB Comparison of analytical and numerical solutions}), and
\item to quantitatively oppose the accuracy of the peak frequencies using spectral significance (\ref{SUB Accuracy of peak frequencies}, \ref{SUB Aliasing}) vs. the `simple' DFT amplitude, the Lomb-Scargle Periodogram, PDM, and the DFT amplitude plus improvement of resulting peak frequencies by least-squares fitting.
\end{itemize}

A frequent problem in the context of astronomical observations is that all measurements do not necessarily have the same variance. This may be due to different instrumentation for different subsets of the time series or changing environmental conditions, such as thermal noise or transparency variations in the Earth's atmosphere. Statistical weights are introduced into the spectral significance in order to take appropriate account of the variable quality of measurements within a single dataset (\ref{SUB Statistical weights}).

\section{Amplitude PDF for non-equidistant sampling}\label{S Amplitude PDF for non-equidistant sampling}

\begin{figure}\includegraphics[width=256pt]{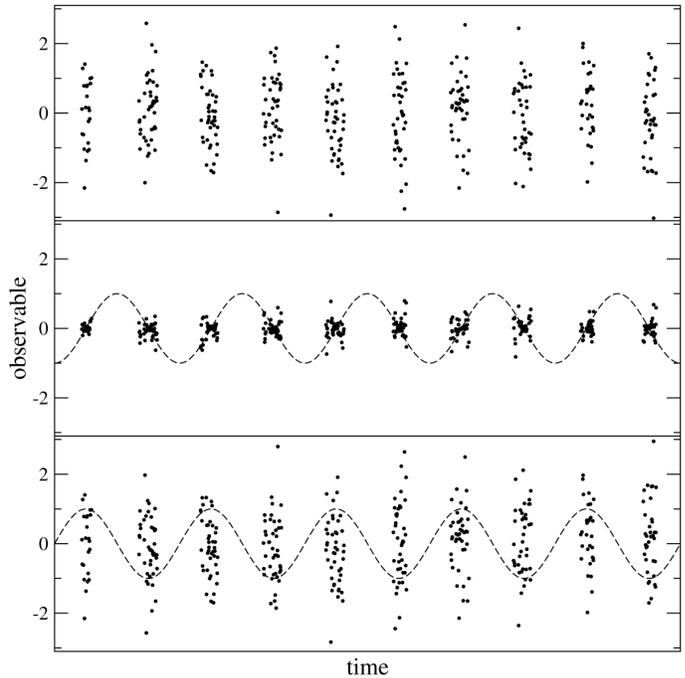}
\caption{Schematic illustration of a time series with periodic gaps representing Gaussian noise with a standard deviation equal 1 ({\em top}) to visualize the dependency of the corresponding frequency-domain noise on frequency and phase, as well as on the characteristics of the time-domain sampling. In the two examples referred to in the lower panels, a test signal is displayed ({\em dashed lines}) for which the DFT shall be evaluated. If the frequency and phase are combined such that the data points consistently align around the times when the trigonometric function attains a small value, only a small fraction of the time-domain noise will be transformed into the frequency domain ({\em mid}). For a combination of frequency and phase grouping the data points around the maxima and minima of the test signal, the time-domain noise will produce a higher frequency-domain noise, correspondingly ({\em bottom}). The first of these two cases produces a narrower probability distribution in Fourier Space, and consequently, a signal with the same amplitude will be considered more reliable in the first case than in the second.}\label{FIGschematic}
\end{figure}

This section presents the theoretical evaluation of the frequency-domain PDF for a non-equidistantly sampled time series representing Gaussian noise.

A very important fact for the statistical analysis is, that in most practical applications, the mean value of the observable is shifted to a fixed value (frequently zero) before the DFT is evaluated. The statistical description of frequency-domain noise has to take this fact into account.

In statistical terms, the Fourier Coefficients are regarded as weighted sums of random variables. Such sums tend towards a Gaussian distribution as datasets become large enough. Since the Fourier Coefficients for each frequency represent the Cartesian components of the two-dimensional Fourier Space, the resulting Gaussian distribution will be two-dimensional (bivariate) as well. Furthermore, as the Fourier Coefficients are functions of frequency, the considered Fourier-Space probability distribution will depend on frequency, correspondingly.

For a time series with gaps\footnote{The time base represents 10 nights of ground-based photometry of the star IC\,4996\,\#\,89 (Zwintz et al.~2004; Zwintz \& Weiss~2006). See \ref{SUBSUB The Sock Diagram} for further detail.} (Fig.\,\ref{FIGschematic}), the relevant influence of the time-domain sampling on the frequency-domain probability distribution is determined by the phase coverage of the measurements: a combination of frequency and phase for which the intervals containing data are consistently associated to angles where the corresponding trigonometric function (`test signal') attains low numerical values will result in a lower noise level than a combination allocating the data close to angles where the maxima/minima of the test signal are located. The first case will yield a narrow probability distribution, the second case a broad one. Consequently, a signal with the same amplitude has be considered more significant in the first case. These strong phase dependencies are mitigated for frequencies providing a better phase coverage.

Consequently, the amplitude distribution in Fourier Space will have to be a function of both {\em frequency and phase}, which is achieved by expressing the bivariate Gaussian PDF in polar rather than Cartesian coordinates.

The probability of a peak generated by noise to reach a given amplitude level may be evaluated through integration of the PDF over amplitude, which leads to the {\em Cumulative Distribution Function (CDF)} and False-Alarm Probability, based on which a more  informative quantity, the spectral significance, is defined.

\subsection{Zero-mean correction}\label{SUB Zero-mean correction}

In astronomical applications, magnitudes are usually averaged to a pre-defined constant (zero or non-zero, as obtained by some theoretical concept or calibration) before the amplitude spectrum is evaluated. The following considerations apply to observables adjusted to zero mean. If a non-zero constant is chosen instead, the DFT will change, but the False-Alarm Probabilities will remain the same. In this case, one prefers to evaluate a DFT spectrum for the time series as it is but use a zero-mean corrected version of the dataset for the computation of spectral significances.

Considering a time series $x_k := x\left( t_k\right)$ to be generated by a Gaussian random process with expected value $0$ and population variance $\left< x^2\right>$, the time-domain PDF is
\begin{equation}
\phi\left( x_k\right) := \frac{1}{\sqrt{2\pi\left< x^2\right>}}\:\mathrm{e}^{-\frac{x_k^2}{2\left< x^2\right>}}\: .
\end{equation}
Given a random process that produces an infinite population of Gaussian random variables, the mean of a finite sample of random variables $x_k$ is free to scatter around the population mean. Gauss's law of error propagation returns a variance of the sample mean which is proportional to the inverse number of data points in the sample,
\begin{equation}
\left<\left< x_k\right> ^2\right> = \frac{\left< x^2\right>}{K}\: .
\end{equation}
If the finite sample is artificially adjusted to zero average, the sample mean value is not allowed to scatter any more. Since the standard error of an individual data point implicitly contains the standard error of the mean, too, zero-mean correction will distort the PDF of the random variable. The only exception with an invariant PDF is the Fourier Analysis, i.\,e., equidistant time-domain sampling and a set of discrete frequencies.

An alternative (and promising) method in this context is the {\em Floating-Mean Periodogram} (Cumming, Marcy \& Butler 1999), which is based on a least-squares fit of a sinusoid {\em plus} a constant to the time series, the latter retaining the free scatter of the sample mean.

\subsection{Distribution of Fourier Coefficients}\label{SUB Distribution of Fourier Coefficients}

Incorporating the effect of zero-mean correction into the statistical examination, the zero-mean corrected magnitude values $x_k - \frac{1}{K}\sum_{k=0}^{K-1}x_k$ have to be used for the calculation of Fourier Coefficients according to\footnote{Some applications prefer a different normalization of eqs.\,\ref{EQ zero-mean a}, \ref{EQ zero-mean b}. E.\,g., in publications dealing with the theoretical aspects of functional analysis, both Fourier Coefficients and the inverse transform from the frequency into the time domain are frequently normalized by $K^{-\frac{1}{2}}$ instead of $K^{-1}$. Also in the field of communications engineering, different normalizations are used. In fact, one is free to distribute the normalization factors among these relations arbitrarily, as long as the product of both factors is $K^{-1}$.}
\begin{eqnarray}
\label{EQ a_ZM}a_{\mathrm{ZM}}\left(\omega\right) & := & \frac{1}{K}\sum_{k=0}^{K-1}x_k\cos\omega t_k - \frac{1}{K^2}\sum_{k=0}^{K-1}x_k\sum_{l=0}^{K-1}\cos\omega t_l \: ,\\
\label{EQ b_ZM}b_{\mathrm{ZM}}\left(\omega\right) & := & \frac{1}{K}\sum_{k=0}^{K-1}x_k\sin\omega t_k - \frac{1}{K^2}\sum_{k=0}^{K-1}x_k\sum_{l=0}^{K-1}\sin\omega t_l \: .
\end{eqnarray}
Due to the linearity of the Fourier Transform, the subtraction of a constant in the time domain refers to a subtraction of a spectral window in the frequency domain\footnote{Since the Fourier Analysis of equidistant time series is restricted to discrete frequencies associated to orthogonal DFTs, nothing will change for non-zero frequencies in this special case.}.

Rearrangement of indices yields
\begin{eqnarray}
\label{EQ zero-mean a}a_\mathrm{ZM}\left(\omega\right) & = & \frac{1}{K}\,\sum_{k=0}^{K-1}x_k\left(\cos\omega t_k - \frac{1}{K}\,\sum_{l=0}^{K-1}\cos\omega t_l\right)\: ,\\
\label{EQ zero-mean b}b_\mathrm{ZM}\left(\omega\right) & = & \frac{1}{K}\,\sum_{k=0}^{K-1}x_k\left(\sin\omega t_k - \frac{1}{K}\,\sum_{l=0}^{K-1}\sin\omega t_l\right)\: .
\end{eqnarray}

Given pure Gaussian noise in the time domain with a population variance $\left< x^2\right>$, eqs.\,\ref{EQ zero-mean a}, \ref{EQ zero-mean b} allows one to consider both Fourier Coefficients as linear combinations of Gaussian variables with expected values $0$ and variances
\begin{eqnarray}
\label{EQ CosCoef variance}\left< a_\mathrm{ZM}^2\right>\left(\omega\right) & = & \frac{\left< x^2\right>}{K^2}\,\sum_{k=0}^{K-1}\left(\cos\omega t_k - \frac{1}{K}\,\sum_{l=0}^{K-1}\cos\omega t_l\right) ^2 ,\\
\label{EQ SinCoef variance}\left< b_\mathrm{ZM}^2\right>\left(\omega\right) & = & \frac{\left< x^2\right>}{K^2}\,\sum_{k=0}^{K-1}\left(\sin\omega t_k - \frac{1}{K}\,\sum_{l=0}^{K-1}\sin\omega t_l\right) ^2 .
\end{eqnarray}

Thanks to the Central Limit Theorem (de Moivre 1718; Stuart \& Ord 1994, pp.\,310f), the consideration of these coefficients as Gaussian variables holds to a sufficient degree, even if the time-domain noise is not Gaussian, because even short datasets in astronomical applications are long enough compared to the fast convergence of the PDF towards the Gaussian distribution with an increasing number of random variables.

\subsection{Frequency- and phase-dependent PDF}\label{SUB Frequency- and phase-dependent PDF}

Since the DFT produces a two-dimensional vector $\left( a,b\right)$, the probability distribution in the frequency domain will also be two-dimensional, so-called bivariate.

The combined probability density of two independent Gaussian variables $\alpha$, $\beta$ with corresponding variances $\left<\alpha ^2\right>$, $\left<\beta ^2\right>$ is given by a bivariate Gaussian PDF,
\begin{equation}\label{EQ ell Gauss}
\phi\left(\alpha ,\beta\right) = \frac{1}{2\pi\sqrt{\left<\alpha ^2\right>\left<\beta ^2\right>}}\,\mathrm{e}^{-\frac{1}{2}\left(\frac{\alpha ^2}{\left<\alpha ^2\right>}+\frac{\beta ^2}{\left<\beta ^2\right>}\right)}\: ,
\end{equation}
if the covariance $\left<\alpha\beta\right>$ vanishes.

The fact that the Fourier Coefficients $a_\mathrm{ZM}$, $b_\mathrm{ZM}$ of pure noise are two linear combinations of the same random vector in Fourier Space diminishes the degrees of freedom by $1$. Hence they may be considered independent to a sufficient degree, if the sample size is large enough. Consequently, if $\left< a_\mathrm{ZM}b_\mathrm{ZM}\right> = 0$, eq.\,\ref{EQ ell Gauss} describes the bivariate distribution of Fourier Coefficients related to noise satisfactorily. According to Appendix \ref{APPENDIX Orientation of the rms Error Ellipse of Fourier Coefficients}, rotating the Fourier Space coordinates by an angle $\theta _0$ given by
\[
\tan 2\theta _0\left(\omega\right) =
\]
\begin{equation}\label{EQ phases of extreme variance}
\frac{K\sum_{k=0}^{K-1}\sin 2\omega t_k-2\sum_{k=0}^{K-1}\cos\omega t_k\sum_{k=0}^{K-1}\sin\omega t_k}{K\sum_{k=0}^{K-1}\cos 2\omega t_k-\left(\sum_{k=0}^{K-1}\cos\omega t_k\right) ^2+\left(\sum_{k=0}^{K-1}\sin\omega t_k\right) ^2}
\end{equation}
transforms the Fourier Coefficients $a_\mathrm{ZM}$, $b_\mathrm{ZM}$ into coefficients $\alpha$, $\beta$ with zero covariance, as desired.

The DFT of a measured time series $x_k$ contains only an amplitude $A$ but also a phase angle
\begin{equation}\label{EQ Fourier phase}
\theta\left(\omega\right) = \frac{\sum_{k=0}^{K-1}x_k\sin{\omega t_k}}{\sum_{k=0}^{K-1}x_k\cos{\omega t_k}}\: .
\end{equation}
This additional information may be taken into account by evaluating the conditional probability density of amplitude for a constant phase angle $\theta$, pre-defined by the DFT of the time series under consideration at the frequency $\omega$.

The transformation of eq.\,\ref{EQ ell Gauss} from Cartesian into polar coordinates $\left( A,\theta\right)$ is performed via $d\,\alpha\,d\,\beta = A\,dA\,d\theta$ (Appendix \ref{SUB General PDF transformation}) with
\begin{eqnarray}
\alpha & = & \frac{A}{2}\cos\left(\theta - \theta _0\right)\: , \\
\beta & = & \frac{A}{2}\sin\left(\theta - \theta _0\right)\: ,
\end{eqnarray}
where the fact, that the coordinate system is rotated by a constant angle $\theta _0$, does not contribute to the Jacobian of the transformation. The division by $2$ is introduced by collecting the contributions of both $A\left(\omega\right)$ and  $A\left( -\omega\right)$ -- which are equal for real observables -- to the total amplitude. The transformed amplitude PDF becomes
\begin{equation}\label{EQ amplitude PDF raw}
\phi\left( A,\theta\left|\right.\omega\right) = \frac{A}{2\pi\sqrt{\left<\alpha ^2\right>\left<\beta ^2\right>}}\,\mathrm{e}^{-\frac{A^2}{8}\left[\frac{\cos ^2\left(\theta -\theta _0\right)}{\left<\alpha ^2\right>}+\frac{\sin ^2\left(\theta -\theta _0\right)}{\left<\beta ^2\right>}\right]}\: .
\end{equation}
Of course, $\phi$ does not only depend on amplitude $A$ and phase $\theta$, but also on $\alpha _0$, $\beta _0$, and $\theta _0$, which are determined by the time-domain sampling and are functions of frequency $\omega$. The bar symbol in $\phi\left( A,\theta\left|\right.\omega\right)$ is introduced to formally separate random variables to those considered constant.

Changing the normalization condition from $\int\!\!\!\int _{\,{\cal R} ^2}dA\,d\theta\,\phi\left( A,\theta\left|\right.\omega\right) = 1$ into $\int _{\,\cal{R}} dA\,\phi\left( A\left|\right.\omega ,\theta\right) = 1$ yields
\begin{equation}\label{EQ amplitude PDF}
\phi \left( A\left|\right.\omega ,\theta\right) = \frac{A}{4R^2}\,\mathrm{e}^{-\frac{A^2}{8R^2}}
\end{equation}
with
\begin{equation}\label{EQ rho}
R := \sqrt{\frac{\left<\alpha ^2\right>\left<\beta ^2\right>}{\left<\beta ^2\right>\cos ^2\left(\theta - \theta _0\right) +\left<\alpha ^2\right>\sin ^2\left(\theta - \theta _0\right)}}\: .
\end{equation}
The difference between eq.\,\ref{EQ amplitude PDF raw} and eq.\,\ref{EQ amplitude PDF} is that eq.\,\ref{EQ amplitude PDF raw} is a bivariate PDF of amplitude and phase, whereas in eq.\,\ref{EQ amplitude PDF} only the amplitude $A$ is considered as a random variable, and $\theta$ is constant. This relation returns the probability density of amplitude for a fixed frequency and a fixed phase in Fourier Space. Accordingly, the PDF is normalized by the condition that its integral over the entire amplitude range (from $0$ to $\infty$) has to be $1$. Furthermore, amplitudes being defined $\ge 0$ introduce a factor $2$ into the argument of the exponential function.

Eq.\,\ref{EQ rho} presents $R\left(\theta\right)$ as an ellipse in polar coordinates. The semi-major and semi-minor axes are $\sqrt{\left<\alpha ^2\right>}$ (eq.\,\ref{EQ CosCoef extreme variance}) and $\sqrt{\left<\beta ^2\right>}$ (eq.\,\ref{EQ SinCoef extreme variance}), respectively, and the orientation is determined by $\theta _0$. This ellipse will be called the {\em rms error ellipse}. Its orientation and dimensions depend on frequency.

Eq.\,\ref{EQ phases of extreme variance} has got a set of solutions for $\theta _0$ assigned to orthogonal directions: if $\theta _0$ is a solution, then the complete set of solutions is $\theta _0 + \frac{z}{2}\pi$ $\forall z\in\cal{Z}$. Whether $\sqrt{\left<\alpha ^2\right>}$ returns the semi-major axis and $\sqrt{\left<\beta ^2\right>}$ the semi-minor axis of the rms error ellipse, or vice versa, depends on the choice of $\theta _0$. This paper consistently uses solutions of $\theta _0$ that assign $\alpha _0$ to semi-minor axes, which yields the maximum spectral significance for all phase angles under consideration.

As shown later (see \ref{SUBSUB The Sock Diagram}), the introduction of the normalized semi-major and semi-minor axes,
\begin{eqnarray}
\nonumber &&\alpha _0\left(\omega ,\theta _0\right) := \sqrt{2K\frac{\left<\alpha ^2\right>}{\left< x^2\right>}} =\\
\label{EQ normalized semi-major axis}&&\sqrt{\frac{2}{K^2}\left\lbrace K\sum_{k=0}^{K-1}\cos ^2\left(\omega t_k - \theta _0\right) - \left[\sum_{l=0}^{K-1}\cos\left(\omega t_l - \theta _0\right)\right] ^2\right\rbrace}\: ,\\
\nonumber &&\beta _0\left(\omega ,\theta _0\right) := \sqrt{2K\frac{\left<\beta ^2\right>}{\left< x^2\right>}} =\\
\label{EQ normalized semi-minor axis}&&\sqrt{\frac{2}{K^2}\left\lbrace K\sum_{k=0}^{K-1}\sin ^2\left(\omega t_k - \theta _0\right) - \left[\sum_{l=0}^{K-1}\sin\left(\omega t_l - \theta _0\right)\right] ^2\right\rbrace}\: ,
\end{eqnarray}
respectively, provides the separation of sampling-dependent quantities from quantities that only depend on the DFT amplitude. In this context, the term `normalized' means that for arguments $\omega t_k - \theta _0$ uniformly distributed on $\left[ 0,2\pi\right]$, the expected values of both $\alpha _0$ and $\beta _0$ are $1$.

Given the orientation of axes, $\theta _0$, and the normalized axes, $\alpha _0$, $\beta _0$, of the ellipse, the standard deviation for an arbitrary phase $\theta$ in Fourier Space is the radius of an ellipse in polar coordinates ($R$, $\theta$) according to 
\begin{equation}\label{EQ rho norm}
R = \sqrt{\frac{\left< x^2\right>}{2K}\:\frac{\alpha _0^2\,\beta _0^2}{\beta _0^2\cos ^2\left(\theta - \theta _0\right) +\alpha _0^2\sin ^2\left(\theta - \theta _0\right)}}\: .
\end{equation}

The three parameters $\theta _0$, $\alpha _0$, $\beta _0$ describe the ellipticity of the two-di\-mensional PDF for Gaussian noise in Fourier Space and thus represent the cornerstones of spectral significance evaluation. All subsequent relations will be given in terms of these three quantities.

\subsection{False-Alarm Probability}\label{SUB CDF and False-Alarm Probability}

The {\em Cumulative Distribution Function (CDF)} is obtained by integrating the PDF (eq.\,\ref{EQ amplitude PDF}) according to
\begin{equation}
\Phi \left( A\left|\right.\omega ,\theta\right) = \int _0^A dA^\prime\phi\left( A^\prime\left|\right.\omega ,\theta\right)\: ,
\end{equation}
which yields
\begin{equation}\label{EQ amplitude CDF}
\Phi \left( A\left|\right.\omega ,\theta\right) = 1 - \mathrm{e}^{-\frac{A^2}{8R^2}}\: .
\end{equation}
Thus the probability for an amplitude to exceed a given limit $A$ is given by
\begin{equation}\label{EQ False-Alarm Probability}
\Phi _\mathrm{FA}\left( A\left|\right.\omega ,\theta\right) = \mathrm{e}^{-\frac{A^2}{8R^2}}\: ,
\end{equation}
which is the False-Alarm Probability of an amplitude level at phase $\theta$ (and frequency $\omega$, since $\theta$, $\theta _0$, $\alpha _0$, and $\beta _0$ are frequency-dependent quantities).

\subsection{Spectral significance}\label{SUB Significance}

The frequency- and phase-dependent False-Alarm Probability of an amplitude level was introduced as the probability that random noise in the time domain with the same rms error as the given time series produces a peak in the DFT amplitude spectrum which is at least as high as the corresponding amplitude level for the time series itself: if a peak is assigned a False-Alarm Probability of $0.00\,001$, its risk of being due to noise is $1 : 100\,000$. In this section, the spectral significance\footnote{to be distinguished from significance in the sense of a confidence threshold as used in hypothesis testing.} as a more informative quantity is introduced (\ref{SUBSUB Definition}). It is the inverse False-Alarm Probability (in this case, $100\,000$) scaled logarithmically. In the present example, the conversion of a False-Alarm Probability of $0.00\,001$ into spectral significance returns $5$. In this context, the spectral significance is presented as a logarithmic measure for the number of cases in one out of which the considered amplitude would be an artifact.

Plotting the spectral significance vs.~frequency yields the significance spectrum, and the identification and consideration of the highest peak in this spectrum may lead to a statement on the significance of the entire spectrum. The argument is similar to many existing significance estimates (e.\,g., Scargle 1982): if the highest peak in the spectrum is below some limit, the entire spectrum has to be considered insignificant. But instead of using the (statistically biased) signal-to-noise ratio as a threshold, the spectral significance is employed. The application of the spectral significance concept to the highest peak out of a sample is briefly discussed (\ref{SUBSUB Spectral significance for a statistically independent sample}).

The formal correspondence to traditional techniques, namely signal-to-noise ratio and the Lomb-Scargle Periodogram, is of special interest and hence provided subsequently (\ref{SUBSUB Connection between significance and signal-to-noise ratio}, \ref{SUBSUB Significance and Lomb Periodogram}).

\subsubsection{Definition}\label{SUBSUB Definition}

To enhance the compatibility to the popular signal-to-noise ratio criterion (see \ref{SUBSUB Connection between significance and signal-to-noise ratio}), the spectral significance of a DFT amplitude is defined as
\begin{equation}\label{EQ significance}
\mathrm{sig}\left( A\left|\right.\omega ,\theta\right) := -\log\Phi _\mathrm{FA}\left( A\left|\right.\omega ,\theta\right)\: ,
\end{equation}
or -- using eqs.\,\ref{EQ rho norm} and \ref{EQ False-Alarm Probability} --
\begin{equation}\label{EQ significance full}
\mathrm{sig}\left( A\left|\right.\omega ,\theta\right) = \frac{KA^2\log\mathrm{e}}{4\left< x^2\right>}\left[\frac{\cos ^2\left(\theta - \theta _0\right)}{\alpha _0^2} + \frac{\sin ^2\left(\theta - \theta _0\right)}{\beta _0^2}\right]\: ,
\end{equation}
with the normalized axes, $\alpha _0$ and $\beta _0$, as defined by eqs.\,\ref{EQ normalized semi-major axis}, \ref{EQ normalized semi-minor axis}, and the orientation of the ellipse in Fourier Space according to eq.\,\ref{EQ phases of extreme variance}. Since the angle $\theta _0$ was chosen to refer to $\alpha _0$ as the semi-minor axis of the rms error ellipse, $\alpha _0$ now corresponds to the phase of maximum spectral significance for a given frequency.

The concept of spectral significance computation relies on the analytical comparison of the DFT amplitude generated by a measured time series to noise at the same variance as the time series under consideration. Unless the population variance $\left< x^2\right>$ of the noise used for comparison is given by theory and/or other observations than the ones under consideration, the sample variance $\left< x_k^2\right>$ of the observable may be used as an estimator.

The Cartesian representation of eq.\,\ref{EQ significance full},
\[
\mathrm{sig}\left( a_{\mathrm{ZM}},b_{\mathrm{ZM}}\left|\right.\omega\right) = \frac{K\log\mathrm{e}}{\left< x^2\right>}\left[\left(\frac{a_{\mathrm{ZM}}\cos\theta _0+b_{\mathrm{ZM}}\sin\theta _0}{\alpha _0}\right) ^2\right.
\]
\begin{equation}\label{EQ significance full cartesian}
\left. + \left(\frac{a_{\mathrm{ZM}}\sin\theta _0-b_{\mathrm{ZM}}\cos\theta _0}{\beta _0}\right) ^2\right]\: ,
\end{equation}
is useful for practical applications and will be employed for the implementation of statistical weights (see \ref{SUB Statistical weights}). The subscript `ZM' indicates zero-mean corrected time series data, in consistency with eqs.\,\ref{EQ a_ZM}, \ref{EQ b_ZM}.

Thanks to the logarithmic scaling, the spectral significance appears as a product form (as opposed to an exponential function), where one factor (the bracket term) contains all information on the ellipticity of the underlying PDF in Fourier Space and is entirely determined by the time-domain sampling. This term is scaled according to the squared amplitude. The serendipitous consequence of this separation is that the evaluation of the bracket term applies to all datasets with the given sampling. In a prewhitening cascade, the sampling of the time series will not change. Consequently, the bracket term remains valid for the entire sequence and has to be computed only once. In the prewhitening cascade itself, it is sufficient to rescale the bracket term by the squared amplitude, which speeds up the computations considerably.

Thanks to this formal separation, it is possible to pack all the characteristics of the time-domain sampling into an amplitude-independent function of frequency and phase. This will lead to the Sock Diagram (\ref{SUBSUB The Sock Diagram}).

A further practical advantage of this separation is the occurrence of the population variance $\left< x^2\right>$ independently of frequency and phase. For small samples, the higher uncertainty of the estimated population variance may be overcome by using the sample variance $\left< x_k^2\right>$ instead of the population variance $\left< x^2\right>$ and increasing the spectral significance limit for peak acceptance accordingly.

\subsubsection{Spectral significance for a statistically independent sample}\label{SUBSUB Spectral significance for a statistically independent sample}

One may desire to evaluate the spectral significance for the highest out of a sample of peaks in the significance spectrum -- in analogy to the procedure presented by Scargle (1982). His arguments may be applied to the spectral significance directly.

For a given spectral significance level $\mathrm{sig}$, the probability of an amplitude level generated by a noise process to exceed the spectral significance limit $\mathrm{sig}$ is the False-Alarm Probability $\Phi _\mathrm{FA}$. It is linked to the spectral significance via eq.\,\ref{EQ significance}. The complementary probability that such an amplitude level is below $\mathrm{sig}$ is $1 - \Phi _\mathrm{FA}$. Given a sample of $N$ such amplitude levels, which are statistically independent, the probability of none exceeding $\mathrm{sig}$ is $\left( 1 - \Phi _\mathrm{FA}\right) ^N$. Again, the complement, $1 - \left( 1 - \Phi _\mathrm{FA}\right) ^N$, returns the probability for at least one peak out of the sample to exceed $\mathrm{sig}$. In other words (and using eq.\,\ref{EQ significance} to substitute for $\Phi _\mathrm{FA}$), the False-Alarm Probability for the maximum of a statistically independent sample of $N$ peaks is
\begin{equation}
\widehat{\Phi _\mathrm{FA}} = 1 - \left( 1 - 10^{-\mathrm{sig}}\right) ^N\: .
\end{equation}
This False-Alarm Probability may be transformed into a spectral significance (using eq.\,\ref{EQ significance} again), which yields
\begin{equation}\label{EQ sigmax}
\widehat{\mathrm{sig}} = -\log\left[ 1 - \left( 1 - 10^{-\mathrm{sig}}\right) ^N\right]
\end{equation}
for the spectral significance of the maximum out of a sample of $N$ statistically independent peaks in the significance spectrum.

Solving eq.\,\ref{EQ sigmax} for $\mathrm{sig}$ yields
\begin{equation}\label{EQ sigmaxinv}
\mathrm{sig} = -\log\left( 1 - \sqrt[N]{1 - 10^{-\widehat{\mathrm{sig}}}}\right)\: ,
\end{equation}
which allows one to immediately convert a chosen threshold for maximum spectral significance into `individual' spectral significance (as given by eq.\,\ref{EQ significance full}). In most practical applications, the approximation
\begin{equation}\label{EQ sigmaxinvapprox}
\mathrm{sig}\approx \widehat{\mathrm{sig}} +\log N
\end{equation}
is sufficiently accurate.

For an equidistant time series consisting of $K$ data points, the number of statistically independent DFT amplitudes is $\frac{K}{2}$, if $K$ is an even number. One may set $N := \frac{K}{2}$ as a rough estimate also for the non-equidistant case, which performs quite reliably in practical applications.

For example, if a maximum spectral significance threshold of $5.46$ shall be applied to a time series consisting of $1\, 000$ data points, the corresponding `individual' spectral significance threshold would be $2.76$, which is in good agreement with the numerical results by Kuschnig et al.~(1997), who obtained a significance of $\approx 3$ in this case, examining $19\, 300$ synthetic time series.

\subsubsection{Connection between spectral significance and signal-to-noise ratio}\label{SUBSUB Connection between significance and signal-to-noise ratio}

The correspondence between spectral significance and signal-to-noise ratio is obtained through substitution of eq.\,\ref{EQ significance full} by the amplitude signal-to-noise ratio $\frac{A}{\left< A\right>}$ according to eq.\,\ref{EQ amplitude noise}. This yields
\[
\mathrm{sig}\left( A\left|\right.\omega ,\theta\right) =
\]
\begin{equation}\label{EQ significance SNR}
\frac{\pi\log\mathrm{e}}{2}\,\left(\frac{A}{\left< A\right>}\right) ^2\left[\frac{\cos ^2\left(\theta - \theta _0\right)}{\alpha _0^2} + \frac{\sin ^2\left(\theta - \theta _0\right)}{\beta _0^2}\right]\: .
\end{equation}
Given uniformly distributed arguments of the trigonometric functions, the expected values of both $\alpha _0$ and $\beta _0$ evaluate to $1$. Thus an approximation for the correspondence between amplitude signal-to-noise ratio and spectral significance is obtained by
\begin{equation}\label{EQ significance SNR approx}
\mathrm{sig}\left( A\right)\approx\frac{\pi\log\mathrm{e}}{4}\,\left(\frac{A}{\left< A\right>}\right) ^2\: .
\end{equation}
For example, an amplitude signal-to-noise ratio of 4 -- as a suggested significance estimator by Breger et al.~(1993) -- roughly corresponds to a spectral significance of
\begin{equation}\label{EQ sig SNR 4}
\mathrm{sig}\left( 4\left< A\right>\right)\approx 4\pi\log\mathrm{e}\approx 5.4575\: .
\end{equation}

A numerical simulation for $42\,597$ time series, each consisting of $14\,400$ equidistantly sampled data points representing a single sinusoidal signal with randomly chosen amplitude, frequency, and phase, plus Gaussian noise with randomly chosen rms error was performed. The agreement with eq.\,\ref{EQ significance SNR approx} for spectral significances $< 200$ and a corresponding signal-to-noise ratio of $\approx 25$ is excellent. At higher spectral significances, the following effect has to be taken into account:

The spectral significance is related to a peak generated by noise with the same variance as the observable under consideration, i.\,e.~{\em without prewhitening}. Frequently, a signal-to-noise ratio calculation relies on the rms {\em residual}, i.\,e.~with the peak under consideration prewhitened. Since (on average) a sinusoidal signal with amplitude $A$ contributes $\frac{A^2}{2}$ to the variance in the time domain, the population variance of residuals in the time domain evaluates to
\begin{equation}
\left< r^2\right> \approx \left< x^2\right> - \frac{A^2}{2}\: .
\end{equation}
Using this relation, eq.\,\ref{EQ amplitude noise} permits one to calculate an amplitude noise level with the considered peak prewhitened, according to
\begin{equation}\label{EQ amplitude noise prewhitened}
N\left( A\right) \approx \sqrt{\frac{\pi}{K}\left(\left< x^2\right> - \frac{A^2}{2}\right)}\: ,
\end{equation}
and writing eq.\,\ref{EQ significance SNR approx} in terms of $N\left( A\right)$ leads to
\begin{equation}\label{EQ significance SNR approx corr}
\mathrm{sig}\left( A\right)\approx\frac{\log\mathrm{e}}{2}\frac{K\pi}{\pi +2K\left[\frac{A}{N\left( A\right)}\right] ^{-2}}\: .
\end{equation}

\begin{figure}\includegraphics[width=256pt]{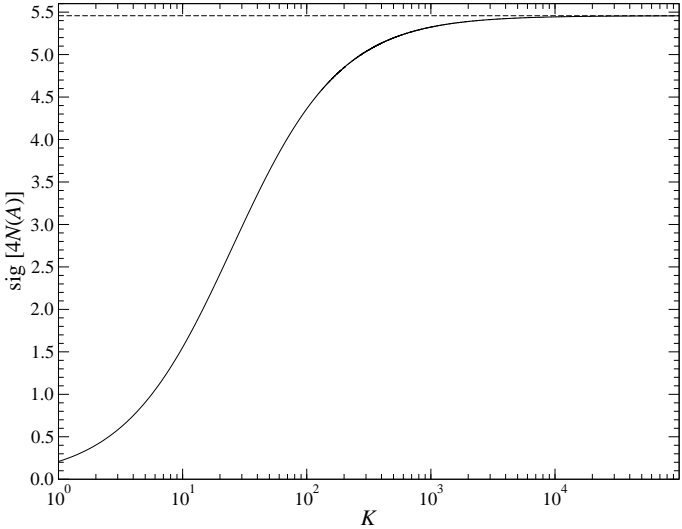}
\caption{Spectral significance (sig) associated with an amplitude signal-to-noise ratio of $4$, without ({\em dashed line}) and with ({\em solid line}) prewhitening of the considered peak, depending on the number of time series data points. If the amplitude noise is calculated after prewhitening, the spectral significance associated with a signal-to-noise ratio of $4$ decreases below the equivalent limit of $5.46$ for short datasets.}\label{FIGsig4pw}
\end{figure}

Fig.\,\ref{FIGsig4pw} shows the difference in spectral significance for amplitude noise with and without prewhitening as a function of time series length. This difference increases dramatically as datasets become shorter. When relying on signal-to-noise ratio estimation, the issue of prewhitening is crucial and in some way philosophical. If the considered peak is believed to be `true' a priori, then it will have to be prewhitened for the noise calculation. If it is assumed to be an artifact, noise will have to be calculated without prewhitening. However, the answer in terms of spectral significance is unique and clear: since the spectral significance analytically compares the considered time series with a randomized one, i.\,e.~keeping the time-domain rms deviation the same, the correspondence to a non-prewhitened signal-to-noise ratio would be correct -- if desired at all. The computation of the spectral significance allows one to completely omit the potentially ambiguous computation of noise levels by averaging amplitude over a frequency interval about a considered peak. In general, the latter are not objective, since the resulting noise level depends on the choice of the interval width and whether unresolved peaks are encountered.

\subsubsection{Spectral significance and Lomb-Scargle Periodogram}\label{SUBSUB Significance and Lomb Periodogram}

If the influence of the zero-mean correction on the statistical characteristics of the Fourier Coefficients is neglected, eqs.\,\ref{EQ normalized semi-major axis}, \ref{EQ normalized semi-minor axis} simplify to
\begin{eqnarray}
\label{EQ alpha0Lomb} \alpha _0\left(\omega ,\theta _0\right) & := & \sqrt{\frac{2}{K^2}\left[ K\sum_{k=0}^{K-1}\cos ^2\left(\omega t_k - \theta _0\right)\right]}\: , \\
\label{EQ beta0Lomb} \beta _0\left(\omega ,\theta _0\right) & := & \sqrt{\frac{2}{K^2}\left[ K\sum_{k=0}^{K-1}\sin ^2\left(\omega t_k - \theta _0\right)\right]}\: ,
\end{eqnarray}
respectively. In this case, the evaluation of $\theta _0$ -- as performed in Appendix\,\ref{APPENDIX Orientation of the rms Error Ellipse of Fourier Coefficients} -- transforms eq.\,\ref{EQ phases of extreme variance} into
\begin{equation}
\label{EQ theta0Lomb} \tan 2\theta _0\left(\omega\right) = \frac{\sum_{k=0}^{K-1} \sin 2\omega t_k}{\sum_{k=0}^{K-1} \cos 2\omega t_k}\: ,
\end{equation}
and eq.\,\ref{EQ significance full} becomes fully compatible to the definition of the Lomb-Scargle Periodogram\footnote{Lomb's (1976) original publication deals with periodogram analysis in terms of power. The expression of the present results in terms of squared amplitude produces explicit consistency.}. In this context, the improvement of the spectral significance compared to the Lomb-Scargle Periodogram is the correct statistical implementation of the artificially fixed time series mean. Remembering that the initial condition that led to the Lomb-Scargle Periodogram was a least-squares solution (unfortunately without handling the zero-mean correction appropriately), the significance spectrum will satisfy the corresponding least-squares condition for the zero-mean corrected data as well. This is in perfect agreement with the results of simulations, as presented in \ref{SUB Accuracy of peak frequencies}.

Since the Fourier-Space effect of the zero-mean correction tends to vanish for infinitely long time series, one would expect the difference between the Lomb-Scargle Periodogram and the spectral significance to become small (or even negligible) for long datasets. The performance of DFT, Lomb-Scargle Periodogram, and spectral significance is compared for time series of different length in \ref{SUB Sample size effects}.

On the other hand, the spectral significance for the Floating-Mean Periodogram (Cumming, Marcy \& Butler 1999), the statistic of which appears not to suffer from zero-mean correction problems, is directly obtained by using eq.\,\ref{EQ significance full} with $\alpha _0$, $\beta _0$ and $\theta _0$ as given above (eqs.\,\ref{EQ alpha0Lomb}, \ref{EQ beta0Lomb}, \ref{EQ theta0Lomb}).

\subsection{The Sock Diagram}\label{SUBSUB The Sock Diagram}

In \ref{SUBSUB Definition}, the spectral significance was introduced as a representation of the statistical properties of time-domain sampling in Fourier Space, applied to the amplitude spectrum of a given observable (eq.\,\ref{EQ significance full}). Selecting a single frequency and phase angle, one finds the spectral significance to be proportional to the squared amplitude. This property makes it easy to determine an analogy to the spectral window.

For classical DFT-based methods, the spectral window is frequently used to determine the effects of time series sampling in the frequency domain. It is defined as the DFT amplitude spectrum of a constant in the time domain, normalized to an amplitude of $1$ at zero frequency. In the case of non-equidistant sampling, peaks in the amplitude window indicate periodicities in the sampling of the time series corresponding to the frequencies where these peaks occur. A frequently returned signature in the spectral window of astronomical single-site measurements is a set of peaks at integer multiples of $1$\,d$^{-1}$. This is the Fourier-Space representation of periodic data gaps due to daylight and termed `$1$\,d$^{-1}$ aliasing'.

Since the spectral significance is a more subtle and sensitive quantity taking into account more information on the time-domain sampling, it is possible to introduce a more sensitive analogy to the spectral window, the {\em Sock Diagram}\footnote{The nomenclature is motivated by the shape of the diagrams, if the time-domain sampling is close to equidistant (see, e.\,g., Fig.\,\ref{FIGzetOphSock}).}. As with all formalism in terms of spectral significance, it is frequency- and phase-resolved.

In analogy to the spectral window, the Sock Diagram represents the spectral significance variations with frequency and phase for a constant amplitude, or amplitude signal-to-noise ratio. It provides quantitative information on the influence by the time-domain sampling on the spectral significance. Furthermore, it displays the quality of signal-to-noise ratio-based estimation for all possible frequencies and phases at a glance. Providing both information on gaps in the sampling and frequency regions with poor accuracy of the DFT amplitude, the Sock Diagram shows where DFT-based signal-to-noise ratio estimation fails.

The normalization of the {\em Sock Function},
\begin{equation}\label{EQ Sock Definition}
\mathrm{sock}\left(\omega ,\theta\right) := \left[\frac{\cos ^2\left(\theta - \theta _0\right)}{\alpha _0^2} + \frac{\sin ^2\left(\theta - \theta _0\right)}{\beta _0^2}\right]\: ,
\end{equation}
provides an expected value $1$ on the assumption of uniformly distributed arguments of the trigonometric functions. Thus
\begin{equation}
\mathrm{sig}\left( A\left|\right.\omega ,\theta\right) = \frac{KA^2\log\mathrm{e}}{4\left< x^2\right>}\:\mathrm{sock}\left(\omega ,\theta\right)\: ,
\end{equation}
as obtained from eq.\,\ref{EQ significance full}, permits to compute the spectral significance associated with an amplitude $A$ based on the Sock Diagram. In terms of signal-to-noise ratio, eq.\,\ref{EQ significance SNR} evaluates to
\begin{equation}
\mathrm{sig}\left( A\left|\right.\omega ,\theta\right) = \frac{\pi\log\mathrm{e}}{4}\,\left(\frac{A}{\left< A\right>}\right) ^2\,\mathrm{sock}\left(\omega ,\theta\right)\: .
\end{equation}

\begin{figure}\includegraphics[width=256pt]{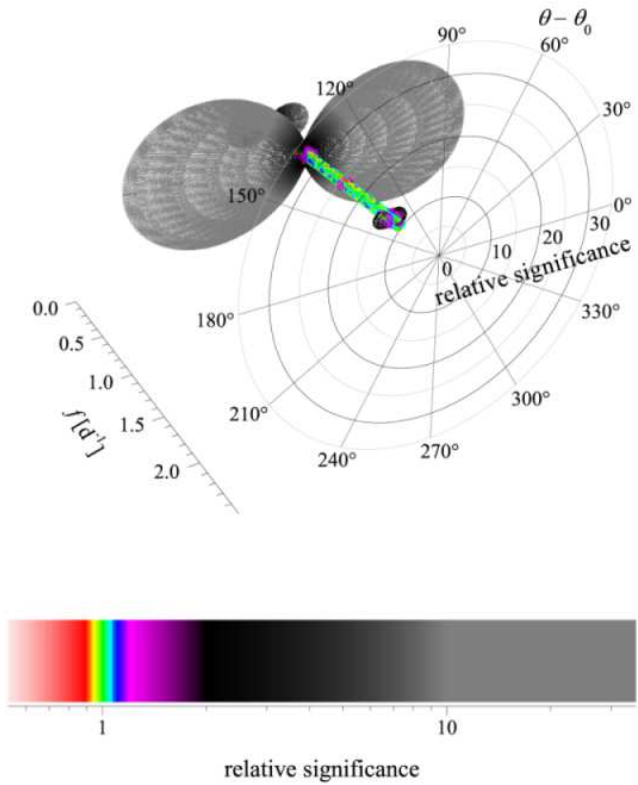}
\caption{Sock Diagram (in cylindrical coordinates) for the $V$ measurements of \object{IC\,4996}\,\#\,89, displaying the relative variations of the spectral significance (radial coordinate) with frequency (height coordinate) and phase (azimuthal coordinate) for a constant signal-to-noise ratio, and hence providing an overview of the effect of time-domain sampling properties in Fourier Space. Better visibility is achieved by additional color coding, referring to the color bar in the lower panel. The susceptibility of DFT spectra to $1$\,d$^{-1}$ aliasing shows up in spectral significance variations of up to a factor $\approx 35$.}\label{FIGsock}
\end{figure}

\begin{figure}\includegraphics[width=256pt]{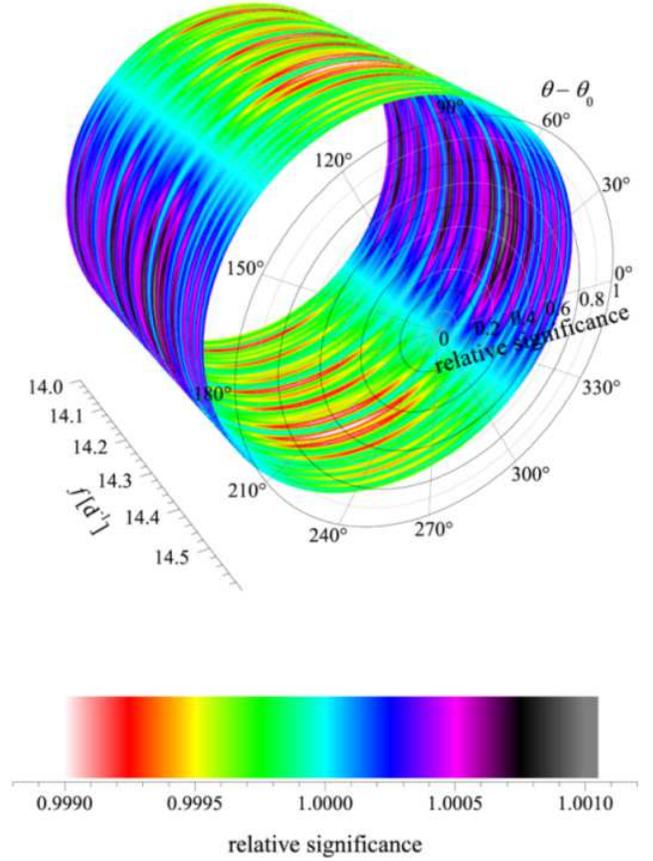}
\caption{Sock Diagram for MOST (`Microvariability and Oscillations of STars'; Walker et al.~2003) measurements of \object{$\zeta$\,Oph}. Due to practically equidistant sampling (duty cycle $\approx 99.9\,\%$, the relative variations of the spectral significance with frequency and phase are $\approx 10^{-3}$, even close to the orbital period of the satellite ($\equiv 14.2$\,d$^{-1}$). As illustrated by the color bar in the lower panel, the color coding scale differs from Fig.\,\ref{FIGsock} considerably.}\label{FIGzetOphSock}
\end{figure}

Since the orientation $\theta _0$ of the elliptical PDF in Fourier Space appears only in the form $\theta - \theta _0$ in all equations related to spectral significance, phase angles in Figs.\,\ref{FIGsock} and \ref{FIGzetOphSock} consistently refer to the position of the semi-minor axis of the rms error ellipse to achieve better visibility: using $\theta - \theta _0$ instead of $\theta$ provides a constant alignment of the spectral significance maxima in phase for all frequencies.

Fig.\,\ref{FIGsock} displays the Sock Diagram for typical non-equidistant sampling representing $10$ nights yielding $381$ data points of single-site $V$ photometry of star \#\,89 in the young open cluster IC\,4996 (Zwintz et al.~2004; Zwintz \& Weiss~2006). Regardless of the astrophysical dimension, these observations have been chosen as the primary test dataset, because they impressively show all the characteristics typical for single-site measurements that make multifrequency analysis a puzzling task.

The Sock Diagram uses three-dimensional polar coordinates: for each frequency, the angular coordinate refers to phase, and the radial component is associated with the spectral significance normalized to an expected value of $1$ (according to eq.\,\ref{EQ Sock Definition}). To enhance the visibility, the radial information is color-coded additionally.

The spectral significance variations close to $0.5$, $1$, $1.5$, and $2$\,d$^{-1}$ clearly indicate frequencies where signal-to-noise ratio estimation of False-Alarm Probability is potentially misleading.

An example for excellent sampling is shown in Fig.\,\ref{FIGzetOphSock}, representing $72\, 055$ MOST\footnote{MOST is a Canadian Space Agency mission, jointly operated by Dynacon Inc., the University of Toronto Institute of Aerospace Studies, the University of British Columbia, and with the assistance of the University of Vienna, Austria.} data points of $\zeta$\,Oph Fabry Imaging photometry (Walker et al.~2004; Walker et al.~2005) with a duty cycle of $\approx 99.9\,\%$, obtained between May\,17 and June\,14, 2004. The data reduction is performed according to the procedure described by Reegen et al.~(2006). Even for frequencies close to $14.2$\,d$^{-1}$ -- corresponding to the orbital period of the satellite ($101.4$\,min; Walker et al.~2003) -- the spectral significance varies with frequency and phase by only $\approx 0.1\,\%$.

\subsection{The marginal distribution of phase angles}\label{SUB The distribution of phase angles}

\begin{figure}\includegraphics[width=256pt]{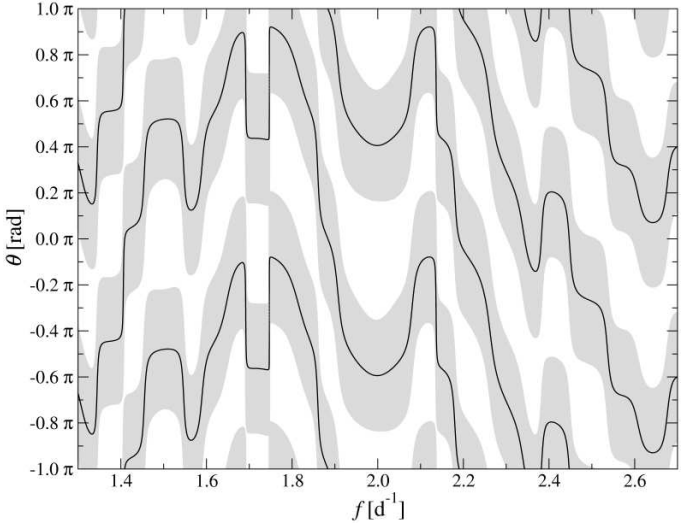}
\caption{Expected values ({\em solid lines}) and rms errors ({\em gray-shaded areas}) of the probability distribution of phase angles in Fourier Space for white noise on the IC\,4996\,\#\,89 sampling.}\label{FIGph}
\end{figure}

An alternative method to the computation of spectral significance as a function of amplitude and phase simultaneously would be to use the phase information separately in addition to, e.\,g., some traditional signal-to-noise ratio-based reliability estimate. The idea is that for white noise in the time domain, the Fourier phases are not uniformly distributed and that the additional incorporation of the phase information may provide a more detailed criterion on the reliability of a peak.

The overall distribution of phases at a given frequency is immediately obtained by integrating the bivariate PDF (eq.\,\ref{EQ amplitude PDF raw}) over amplitude, which yields
\begin{equation}\label{EQ phase PDF}
\phi\left(\theta\right) = \frac{K\alpha _0\,\beta _0}{\beta _0^2\cos ^2\left(\theta -\theta _0\right) - \alpha _0^2\sin ^2\left(\theta -\theta _0\right)}\: ,
\end{equation}
normalized for phases $\theta$ on an interval of width $\pi$. This avoids the ambiguity of solutions for $\theta _0$ modulo $\pi$, as discussed in \ref{SUB Frequency- and phase-dependent PDF}. 

The expected value of this probability distribution is $\theta _0 - \frac{\pi}{2}$: phases associated to a low spectral significance (for given amplitude) generally occur more frequently than phases for which the spectral significance is high. The special case $\alpha _0 = \beta _0$ yields an upper limit of $\frac{\pi}{2\sqrt{3}} \approx 52^{\circ}$ for the standard deviation obtained by eq.\,\ref{EQ phase PDF}.

In statistical terms, the phase distribution provided above is a {\em marginal distribution} of the bivariate PDF given by eq.\,\ref{EQ amplitude PDF raw}. Instead of statistically examining a bivariate distribution, one may use the two marginal distributions (in the present case, the amplitude and phase distributions) instead, but encountering a loss of accuracy, since correlations between the two random variables remain unresolved. Therefore, from the theoretical point of view, it is advisable to use the bivariate form rather than the marginal distributions. An additional problem is that there is no analytical solution for the integral over phase, which would transform the bivariate PDF into the marginal distribution of amplitudes. One would have to employ classical techniques relying on the amplitude signal-to-noise ratio instead.

In addition to these theoretical objections to the examination of marginal distributions, there is a major practical constraint: Fig.\,\ref{FIGph} displays the expected values and standard deviations of Fourier phase for frequencies between 1.3 and 2.7\,d$^{-1}$, given the sampling of IC\,4996\,\#\,89. In the considered frequency range, the rms scatter of phases is greater than $44^{\circ}$ for the entire frequency range under consideration. Compared to the upper limit of $52^{\circ}$, this scatter will probably be much too high to reveal additional information on the reliability of a peak, if the marginal distribution of phases is considered.

The PDF of phases provided by eq.\,\ref{EQ phase PDF} are in perfect agreement to the results of numerical simulations.

\subsection{Spectral significance-based signal recovery}\label{SUB Significance-based signal recovery}

Practical applications are frequently based on a cascade of consecutive prewhitenings. In this case, the frequency at maximum spectral significance, $\hat{\omega}$, is considered as that of the strongest signal component, which is understood to be the next candidate for prewhitening. The next step is to determine the best fit for amplitude and phase at $\hat{\omega}$. The least-squares condition for the best sinusoidal fit at constant frequency $\hat{\omega}$ is obtained by
\begin{eqnarray}
\frac{\partial}{\partial\hat{A}}\sum_{k=0}^{K-1}\left\lbrace x_k -\hat{A}\left[\cos\left(\hat{\omega} t_k-\hat{\theta}\right) -\left<\cos\left(\hat{\omega} t_l-\hat{\theta}\right)\right>\right]\right\rbrace ^2 & = & 0\: , \\
\frac{\partial}{\partial\,\hat{\theta}}\sum_{k=0}^{K-1}\left\lbrace x_k -\hat{A}\left[\cos\left(\hat{\omega} t_k-\hat{\theta}\right) -\left<\cos\left(\hat{\omega} t_l-\hat{\theta}\right)\right>\right]\right\rbrace ^2 & = & 0\: ,
\end{eqnarray}
where the term $\left<\cos\left(\hat{\omega} t_l-\hat{\theta}\right)\right> = \frac{1}{K}\sum_{l=0}^{K-1}\cos\left(\hat{\omega} t_l-\hat{\theta}\right)$ takes into account that the discrete fit has to be zero-mean corrected, if applied to zero-mean corrected data $x_k$. The derivatives lead to
\begin{eqnarray}
\nonumber &&\sum_{k=0}^{K-1} x_k\cos\left(\hat{\omega} t_k -\hat{\theta}\right) = \hat{A}\left\lbrace\sum_{k=0}^{K-1}\cos ^2\left(\hat{\omega} t_k -\hat{\theta}\right)\right. \\
\label{EQ recovery amp condition}&&-\left.\frac{1}{K}\left[\cos\left(\hat{\omega} t_k -\hat{\theta}\right)\right] ^2\right\rbrace\: ,\\
\nonumber &&\sum_{k=0}^{K-1} x_k\sin\left(\hat{\omega} t_k -\hat{\theta}\right) = \hat{A}\left\lbrace\sum_{k=0}^{K-1}\cos\left(\hat{\omega} t_k -\hat{\theta}\right)\sin\left(\hat{\omega} t_k -\hat{\theta}\right)\right.\\
&& - \left.\frac{1}{K}\left[\cos\left(\hat{\omega} t_k -\hat{\theta}\right)\right]\left[\sin\left(\hat{\omega} t_k -\hat{\theta}\right)\right]\right\rbrace\: .
\end{eqnarray}
Solving for $\hat{\theta}$ yields
\[
\sum_{k=0}^{K-1}\sum_{l=0}^{K-1}\sum_{m=0}^{K-1}\left[\sin\hat{\omega}\left( t_l-t_k\right) -\sin\hat{\omega}\left( t_m-t_k\right)\right] x_k\cos\left(\hat{\omega} t_l-\hat{\theta}\right)
\]
\begin{equation}
= 0\: ,
\end{equation}
which finally reduces to
\begin{equation}\label{EQ theta least-squares}
\tan\hat{\theta} = \frac{P_1\sum _{k=0}^{K-1}\cos\hat{\omega}t_k - P_2\sum _{k=0}^{K-1}\sin\hat{\omega}t_k}{P_1\sum _{k=0}^{K-1}\sin\hat{\omega}t_k - P_3\sum _{k=0}^{K-1}\cos\hat{\omega}t_k}\: ,
\end{equation}
using
\begin{eqnarray}
P_1 & := & K\sum _{k=0}^{K-1}\cos\hat{\omega} t_k\sin\hat{\omega} t_k -\left(\sum _{k=0}^{K-1}\cos\hat{\omega} t_k\right)\left(\sum _{k=0}^{K-1}\sin\hat{\omega} t_k\right)\: , \\
P_2 & := & K\sum _{k=0}^{K-1}\cos ^2\hat{\omega} t_k - \left(\sum _{k=0}^{K-1}\cos\hat{\omega} t_k\right) ^2\: , \\
P_3 & := & K\sum _{k=0}^{K-1}\sin ^2\hat{\omega} t_k - \left(\sum _{k=0}^{K-1}\sin\hat{\omega} t_k\right) ^2\: .
\end{eqnarray}
Eq.\,\ref{EQ theta least-squares} provides two solutions for $\hat{\theta}$. To pick the least-squares related solution is an easy task for a program.

Once frequency $\hat{\omega}$ and phase $\hat{\theta}$ of the signal are evaluated, eq.\,\ref{EQ recovery amp condition} immediately yields the best-fitting amplitude,
\begin{equation}
\hat{A} = \frac{\sum_{k=0}^{K-1} x_k\cos\left(\hat{\omega} t_k-\hat{\theta}\right)}{\sum_{k=0}^{K-1}\cos ^2\left(\hat{\omega} t_k-\hat{\theta}\right) - \left[\sum_{k=0}^{K-1}\cos\left(\hat{\omega} t_k-\hat{\theta}\right)\right] ^2}\: .
\end{equation}

\section{Numerical tests}\label{S Numerical tests}

Two sets of numerical simulations have been performed, the first one to confirm the agreement between the theoretically evaluated spectral significance and a straight-forward histogram analysis, and the second one to quantitatively compare the accuracy of frequencies returned by the various methods discussed in this paper.

\subsection{Comparison of analytical and numerical solutions}\label{SUB Comparison of analytical and numerical solutions}

Extensive numerical simulations have been performed in order to examine the validity of the above theoretical considerations. The simulations are set up in a way that compares closely to real life. The algorithm to generate Gaussian noise is based on the Central Limit Theorem (Stuart \& Ord 1994, pp.\,310f), providing fast convergence of a mean value of uniformly distributed random variables towards a Gaussian distribution with increasing number of variables. All of the following numerical applications rely on Gaussian noise produced by summing up $10$ values from the random number generator. A comprehensive compilation on alternative methods to generate Gaussian noise is provided by Firneis (1970).

An all-in-one simulation for many frequencies, many phases, and many amplitudes would result in tremendously time-consuming computations. In order to keep the effort reasonable, only a single frequency is picked to examine the phase dependency of the spectral significance for different amplitudes.

\begin{figure}\includegraphics[width=256pt]{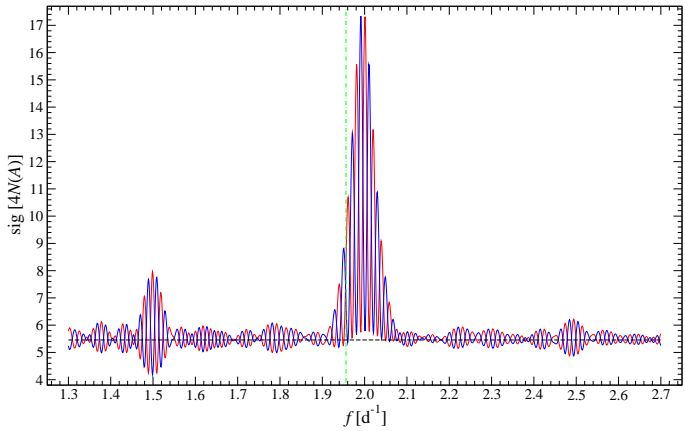}
\caption{Spectral significance associated with an amplitude signal-to-noise ratio of 4 for the $V$ measurements of IC\,4996\,\#\,89. The {\em blue} and {\em red} graphs refer to orthogonal phases in Fourier Space. The expected spectral significance of $\approx 5.46$ is displayed by the {\em dashed black} line. The vertical {\em dashed-dotted green} line indicates the frequency of $1.956$\,d$^{-1}$, which was selected for a numerical simulation to check the validity of the theoretical approach.}\label{FIGFreqProf}
\end{figure}

Fig.\,\ref{FIGFreqProf} displays the spectral significance associated with an amplitude signal-to-noise ratio of 4 -- according to eq.\,\ref{EQ significance SNR} -- for the $V$ data of IC\,4996\,\#\,89. The blue and red graphs represent two orthogonal phases in Fourier Space. The comparison of numerical and theoretical results is performed for a frequency of $1.956$\,d$^{-1}$, where the deviation of spectral significances from the expected value $5.46$ is $\approx 5$ for selected phase angles. Since it is desired to examine the presence of spectral significance variations with phase in numerical results where predicted by theory, this frequency is a reasonable choice for a test.

The procedure consists of five steps.
\begin{enumerate}
\item Zero-mean Gaussian noise with a standard deviation of $1$ is imposed upon the IC\,4996\,\#\,89 sampling.
\item The zero-mean correction is performed to avoid scatter in the mean of the finite time series about the population mean (see \ref{SUB Zero-mean correction}).
\item The Fourier-Space phase angle for the synthetic time series at $1.956$\,d$^{-1}$ is evaluated.
\item The Fourier-Space noise is evaluated upon the variance of the time series according to eq.\,\ref{EQ amplitude noise}.
\item If the Fourier Amplitude exceeds the signal-to-noise ratio under consideration, the resulting histogram is updated correspondingly.
\end{enumerate}
The number of amplitudes exceeding the preselected limit relative to the total number of synthetic datasets provides an estimator for the False-Alarm Probability (and consequently spectral significance) associated with the chosen signal-to-noise ratio.

\begin{figure}\includegraphics[width=256pt]{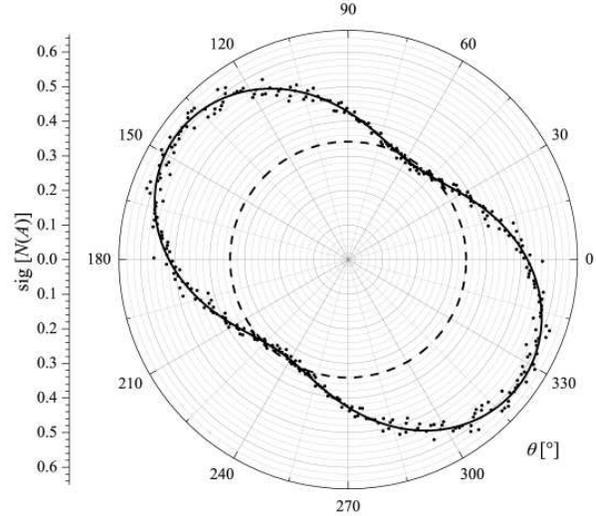}
\caption{Phase-dependent spectral significance (radial coordinate, referred to by the vertical axis) for the $V$ measurements of IC\,4996\,\#\,89 at a constant frequency of $1.956$\,d$^{-1}$ as a function of phase, referring to an amplitude signal-to-noise ratio of 1. The {\em solid line} represents the theoretical result. A numerical simulation for $1.5$ million synthetic datasets ({\em dots}) -- counted in phase bins of width $1^{\circ}$ -- illustrates the excellent agreement. The systematically higher deviation of the numerical results from the theoretical solutions at higher spectral significances is due to sample-size effects: phase bins associated with a high theoretical spectral significance are hit less often than others. Hence the total number of Fourier Amplitudes to be examined differs from phase bin to phase bin. The {\em dashed line} represents the expected spectral significance of $\approx 0.34$ associated with a signal-to-noise ratio of 1 for uniformly distributed arguments of all trigonometric functions.}\label{FIGPhaseProf1}
\end{figure}

\begin{figure}\includegraphics[width=256pt]{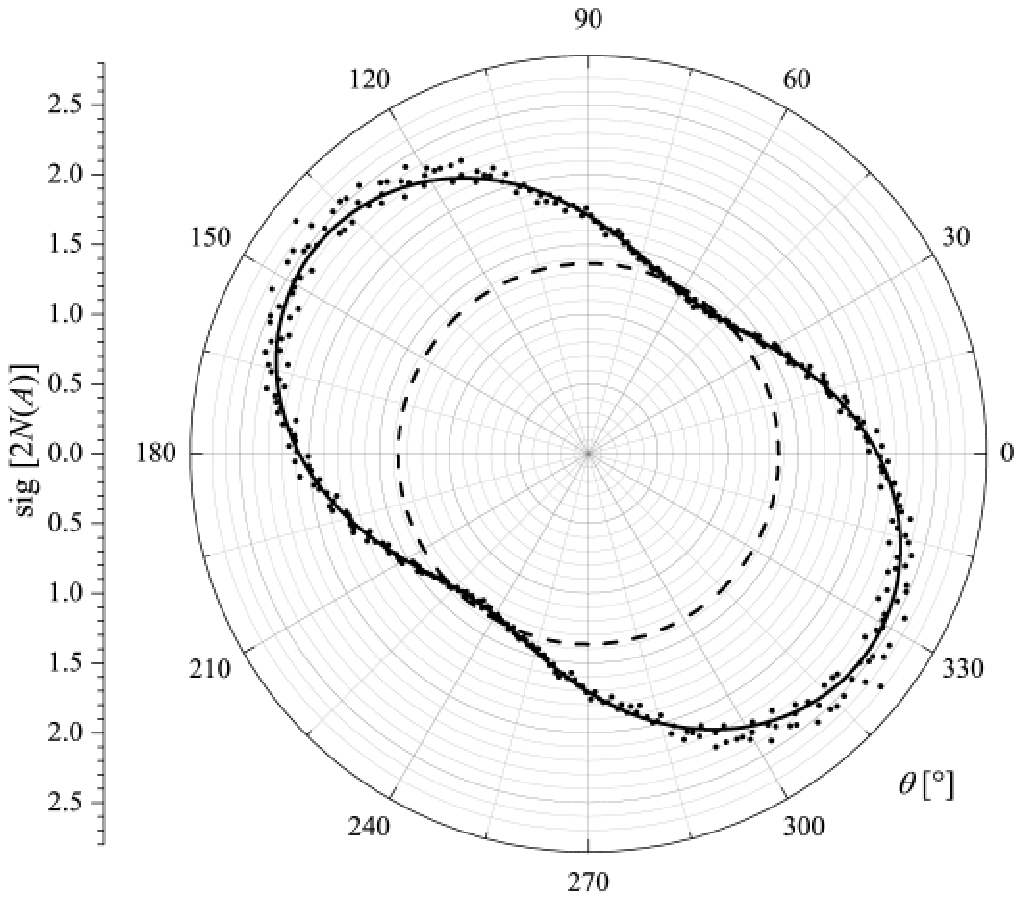}
\caption{Same as Fig.\,\ref{FIGPhaseProf1}, but for an amplitude signal-to-noise ratio of 2. The {\em dots} refer to a numerical simulation for $22$ million synthetic datasets. The expected spectral significance level is $\approx 1.36$ ({\em dashed line}).}\label{FIGPhaseProf2}
\end{figure}

\begin{figure}\includegraphics[width=256pt]{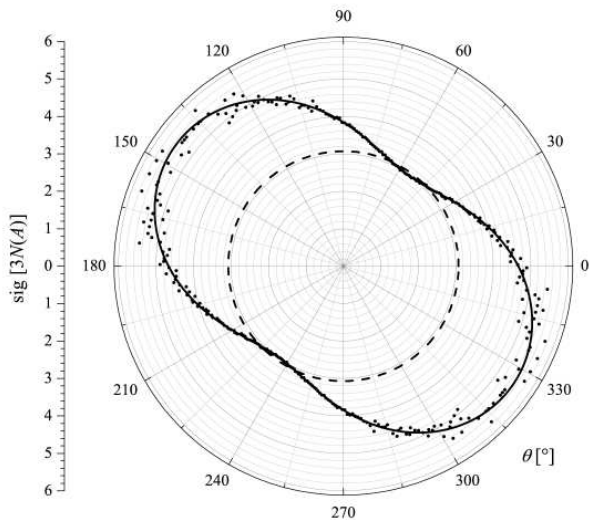}
\caption{Same as Fig.\,\ref{FIGPhaseProf1}, but for an amplitude signal-to-noise ratio of 3. The {\em dots} refer to a numerical simulation for $250$ million synthetic datasets. The expected spectral significance level is $\approx 3.07$ ({\em dashed line}).}\label{FIGPhaseProf3}
\end{figure}

Figs.\,\ref{FIGPhaseProf1} to \ref{FIGPhaseProf3} display the agreement between theory and simulations for amplitude signal-to-noise ratios of 1, 2, and 3. The corresponding numbers of synthetic datasets computed are $1.5$ million, $22$ million, and $250$ million, correspondingly. In all three cases, phase bins of width $1^{\circ}$ were used for the histograms. For a signal-to-noise ratio of $4$, the number of synthetic datasets required for an acceptable numerical accuracy would exceed the capabilities of computational performance, especially those of system random number sequences, by far.

Phases associated with a high spectral significance occur less frequently than others (see \ref{SUB The distribution of phase angles}), whence the scatter of numerical results is systematically higher at higher spectral significance levels due to sample-size effects. Taking this into account, the overall quality of fits is good. In neither of the three plots, a systematic deviation of the numerical results from the analytical functions is visible. However, the theoretical prediction for the overall shape is recovered by the numerical results, indicating that the ellipticity given by the normalized axes $\alpha _0$ and $\beta _0$ matches the `reality' of simulation. Also the orientation of the ellipse ($\theta _0$) appears consistent with the synthetic data. Finally, the consecutive comparison of Figs.\,\ref{FIGPhaseProf1} to \ref{FIGPhaseProf3} offer convincing evidence of the spectral significance at a constant frequency and phase angle to be proportional to the squared amplitude, as predicted by the theoretical solution.

\subsection{Accuracy of peak frequencies}\label{SUB Accuracy of peak frequencies}

One of the major issues in period searches is the accuracy of the resulting signal frequencies. Classical DFT with or without improvement of peak frequencies by a subsequent least-squares fit\footnote{DFT suffers from systematic deviations of peak frequencies (Kovacs~1980).}, Lomb-Scargle Periodogram, and spectral significance analysis are based on the identification of the highest peak in a chosen frequency interval.

A quantitative description of the quality of resulting frequencies with and without noise was obtained by means of simulations. For the sampling of the IC\,4996\,\#\,89 dataset ($V$), the following procedure is performed:
\begin{enumerate}
\item a synthetic signal of given frequency $f_0$ and amplitude with random phase (uniformly distributed on $\left[ -\pi, \pi\right]$) is generated,
\item Gaussian noise with given standard deviation is added (optional),
\item zero-mean correction of the resulting dataset is performed,
\item a DFT amplitude spectrum, a Lomb-Scargle Periodogram, a spectrum of phase dispersion\footnote{The PDM tested here is based on 300 equidistant phase bins of constant width $\frac{\pi}{10}$ rad.}, and a significance spectrum are computed for a predefined frequency range,
\item an interpolation routine is performed to find the maximum (or minimum phase dispersion, respectively) in each of the four spectra, and
\item the deviation of the resulting frequency from the corresponding input frequency, $\Delta f$, is evaluated.
\end{enumerate}

For each frequency, deviations $\Delta f$ were collected to obtain a fre\-quen\-cy-dependent rms error of recovered frequencies, where only attempts with $\left|\Delta f\right| \le 0.5$\,d$^{-1}$ were taken into account. Attempts resulting in $\left|\Delta f\right| > 0.5$\,d$^{-1}$ were considered alias.

\begin{figure*}\includegraphics[width=524pt]{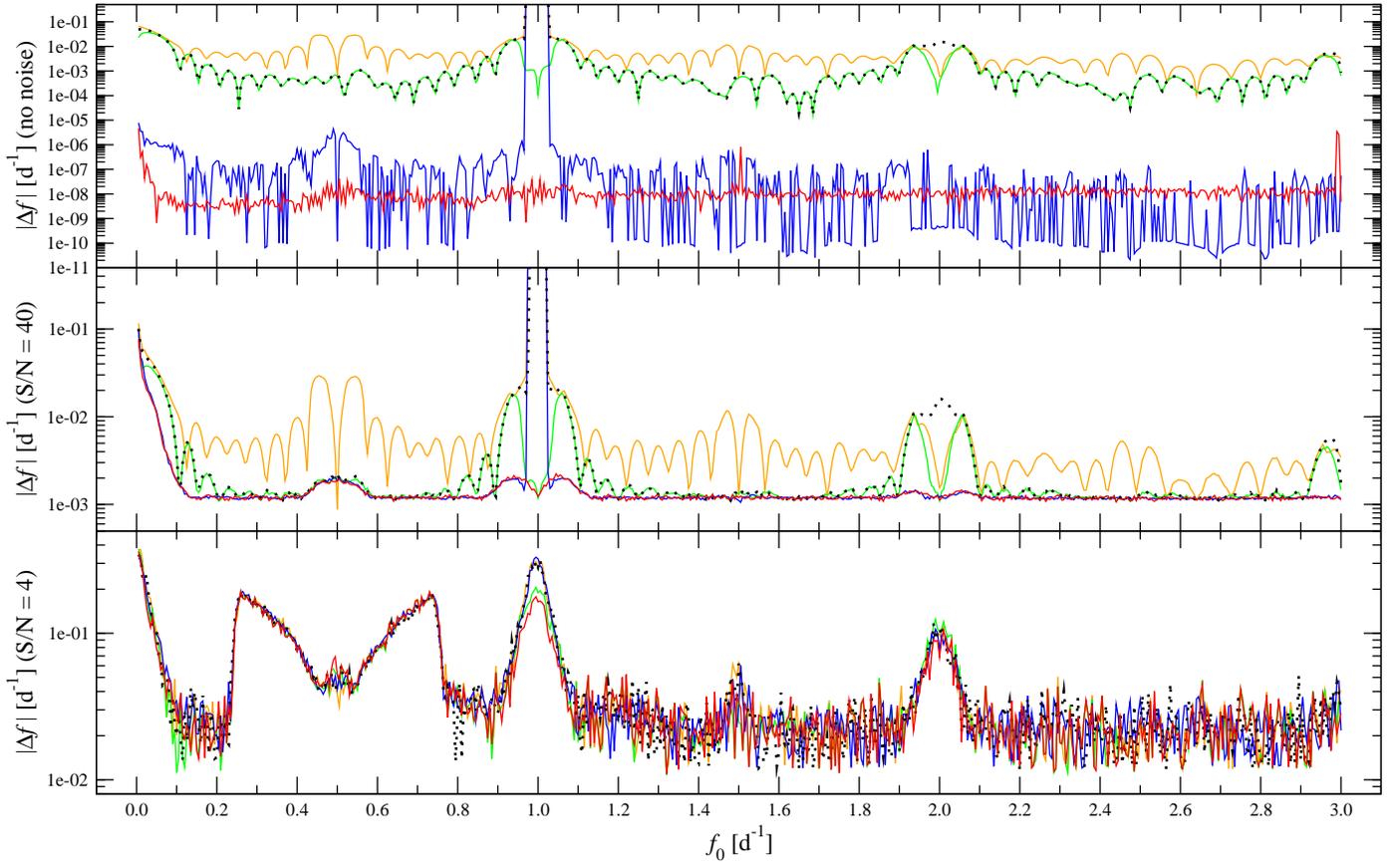}
\caption{Frequency accuracy (rms scatter of resulting frequencies about the initial frequency for a single sinusoidal signal with uniformly distributed phase) vs.~signal frequency of five methods: DFT ({\em solid orange line}), DFT plus least-squares fitting ({\em solid blue line}), Lomb-Scargle Periodogram ({\em solid green line}), PDM ({\em dotted black line}), and spectral significance ({\em solid red line}). The time-domain sampling represents the $V$ measurements of IC\,4996\,\#\,89. The {\em top} panel is the frequency accuracy for a pure signal. Towards the {\em bottom} panel, Gaussian noise with increasing standard deviation is added to the signal in two steps, corresponding to amplitude signal-to-noise ratios of 40 and 4, respectively. Only those attempts where the distance between resulting frequency and input frequency does not exceed $0.5$\,d$^{-1}$ were taken into account, the rest was considered alias (see \ref{SUB Aliasing}). The results are based on a numerical simulation investigating $1\, 000$ datasets for every frequency.}\label{FIGfacc}
\end{figure*}

\begin{figure*}\includegraphics[height=524pt, angle=270]{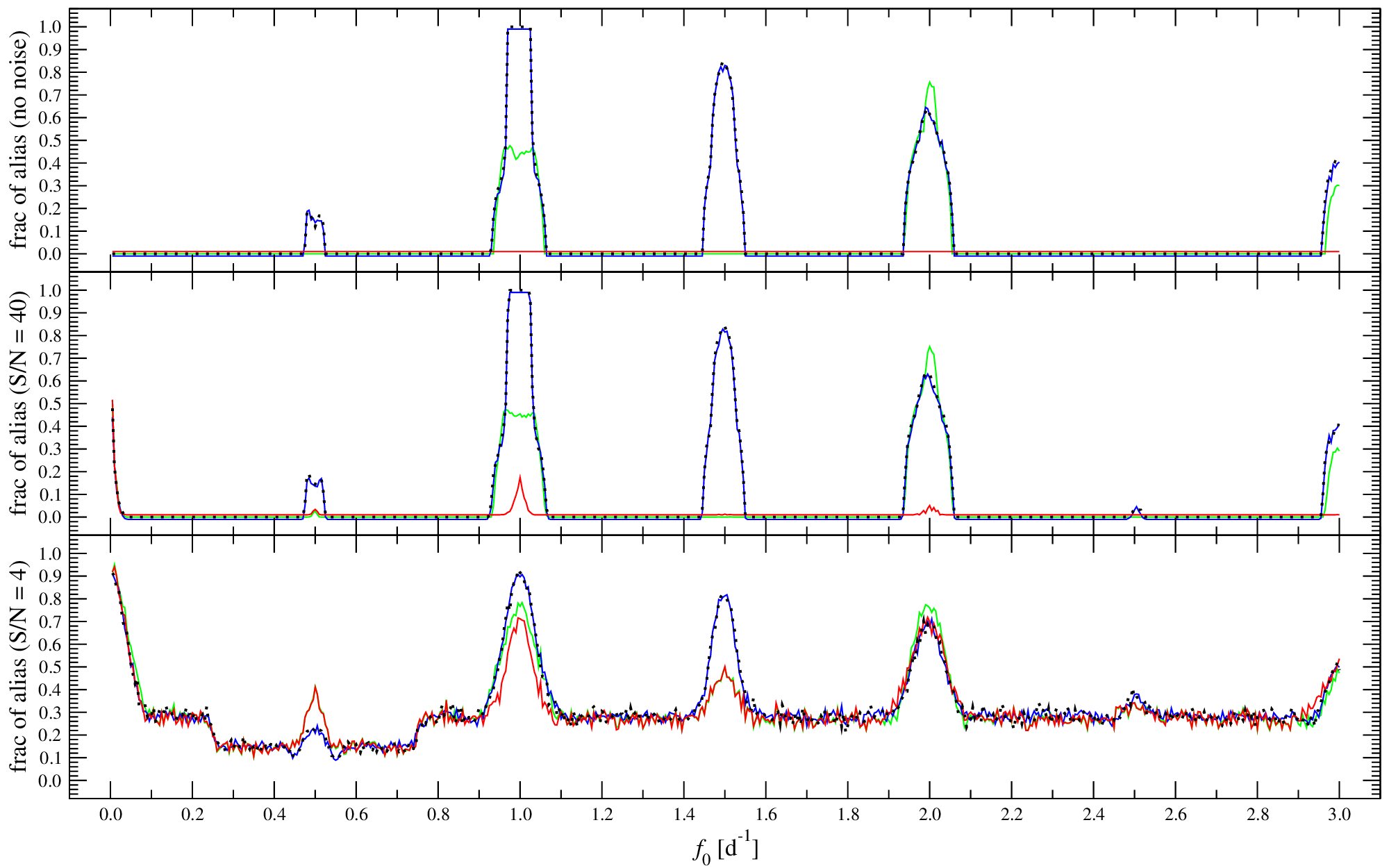}
\caption{Relative number of aliases vs.~signal frequency of four methods: DFT plus least-squares fitting ({\em dashed blue line}), Lomb-Scargle Periodogram ({\em solid green line}), PDM ({\em dotted black line}), and spectral significance ({\em solid red line}). In this context, a result is considered alias, if the absolute difference between resulting frequency and signal frequency exceeds $0.5$\,d$^{-1}$. The DFT without least-squares fitting is not displayed here. It produces essentially the same fraction of alias, because the least-squares fit allows a fine tuning of resulting peak frequencies only, which is negligible considering frequency errors of $0.5$\,d$^{-1}$ and more. The time-domain sampling represents the $V$ measurements of IC\,4996\,\#\,89. The {\em top} panel is the fraction of alias peaks for a pure signal. Towards the {\em bottom} panel, Gaussian noise with increasing standard deviation is added to the signal in two steps, corresponding to amplitude signal-to-noise ratios of 40 and 4, respectively. In the upper two panels, the DFT graph is offset vertically by $-0.01$ and the Lomb-Scargle graph by $+0.01$ to provide better visibility at and close to zero. The results are based on a numerical simulation investigating $1\, 000$ datasets for every frequency.}\label{FIGalias}
\end{figure*}

The example illustrated by Figs.\,\ref{FIGfacc}, \ref{FIGalias} uses the IC\,4996\,\#\,89 data.

Fig.\,\ref{FIGfacc} shows the comparison of the five methods. For each frequency, $1\, 000$ datasets were examined. The figure displays the rms deviation (denoted $\left|\Delta f\right|$) of resulting frequencies as a function of the signal frequency $f_0$. The top panel refers to a clean sinusoidal signal without noise. Towards the bottom panel, Gaussian noise with increasing standard deviation is added (corresponding to signal-to-noise ratios of 40 and 4, respectively, using eq.\,\ref{EQ significance SNR approx corr}).

For a signal without noise (top panel), the frequency accuracies of the Lomb-Scargle Periodogram exceed that of the DFT by a factor $\approx 10$, and both profiles show accuracy variations with frequency. The Lomb-Scargle Periodogram does not take into account zero-mean correction, as shown in \ref{SUBSUB Significance and Lomb Periodogram}. This leads to systematic effects not only for short datasets, but also in frequency regions where the phase coverage becomes poor. In the absence of noise, frequencies at maximum spectral significance are $\approx 100\,000$ times more accurate than the DFT results, which is in the accuracy domain of the least-squares solutions. Apparently the accuracy of the spectral significance peaks is only limited by the internal accuracy of the computer for double-precision floating-point numbers. In addition, the accuracy of spectral significance is, in this respect, practically independent of frequency.

Once noise is added to the signal, the peak frequency accuracy of spectral significance becomes much poorer (in consistency with O'Donoghue \& Montgomery 1999), but as illustrated by Fig.\,\ref{FIGfacc}, the method does not get worse than either alternative procedure. The improvement of the frequency accuracy by the spectral significance solution for high signal-to-noise ratio is valuable, since the exact knowledge of frequency provides exact prewhitening. Only exact prewhitening guarantees that weaker frequency components are not contaminated by spurious residuals of the higher peaks.

\subsection{Aliasing}\label{SUB Aliasing}

One potential weakness of the DFT, which usually becomes puzzling in the analysis of non-equidistantly sampled time series, is the susceptibility to aliasing. Whereas measurements over infinite time would lead to perfectly delta-shaped signal components, the spectral window of `real' data is convolved into this `ideal' spectrum. For typical single-site observations, 1\,d$^{-1}$ sidelobes are visible around every signal peak, and in many cases it is hard to decide which peak out of the `comb' of aliases is the right one. For multiperiodic signal, aliases of individual components may interfere. If this interference leads to an amplification, an erroneous component identification is inevitable, when only relying on the highest amplitude. This applies to the spectral significance as well and reflects a major weakness of the step-by-step prewhitening technique rather than the calculation of the spectral significance. Strategies to overcome the aliasing problem have been examined for many years (e.\,g. Ferraz-Mello 1981; Roberts, Lehar \& Dreher 1987; Foster 1995).

Fig.\,\ref{FIGalias} displays a comparison of DFT, Lomb-Scargle Periodogram, PDM, and spectral significance in terms of fraction of aliases among the $1\, 000$ simulated test datasets used in \ref{SUB Accuracy of peak frequencies}. As in Fig.\,\ref{FIGfacc}, the top panel refers to signal without noise, and Gaussian noise of increasing standard deviation is added towards the bottom panel. Not surprisingly, the susceptibility of the DFT to aliasing does not improve, if a least-squares algorithm is appended. The portion of aliases obtained using the Lomb-Scargle Periodogram is fairly below that of the DFT. Finally, the figure clearly shows that the spectral significance analysis is more stable against aliasing than either compared strategy. Without noise, not a single alias peak occurs among altogether 300\,000 simulated datasets. For a signal-to-noise ratio of $40$, the percentage of alias peaks is less than a third of the corresponding percentage of the alternative methods, even at 1\,d$^{-1}$. Only for the highest noise level, to which the bottom panel refers, all results are quite comparable.

Since the spectral significance is not initially intended to correct for aliasing, its capabilities to avoid potential misidentification of peaks are limited to the extent of spuriously amplified peaks by systematic inhomogeneities of the frequency-domain noise. Simulations and practical experience show that the systematic errors of spectral noise are an inferior error source compared to the interference of aliases of different signal components. Presently investigations are performed to apply the spectral significance technique to simultaneous multi-frequency solving algorithms for an improved treatment of aliases. The results are planned to be presented in paper II.

\subsection{Sample size effects}\label{SUB Sample size effects}

A further example is provided to compare the performance of DFT, Lomb-Scargle Periodogram, and spectral significance for datasets of equal (or at least comparable) characteristics, but different length.

The `long' dataset represents 34 nights (not consecutive, but covering $81$\,d) of single-site photometry of the Delta Scuti star \object{44\,Tau} (Str\o mgren $y$, 2\,981 data points; Antoci et al.~2006) acquired by the Vienna University Automatic Photoelectric Telescope (APT; Strassmeier et al.~1997; Granzer et al.~2001). More details on the data are provided in Section\,\ref{S Practical application}. The `short' dataset is a subset of 7 consecutive nights (619 points). For the subsequent investigations, only the sampling of these two datasets is used. Peak frequency accuracies are computed according to the procedure described in \ref{SUB Accuracy of peak frequencies}.

\begin{figure}\includegraphics[width=256pt]{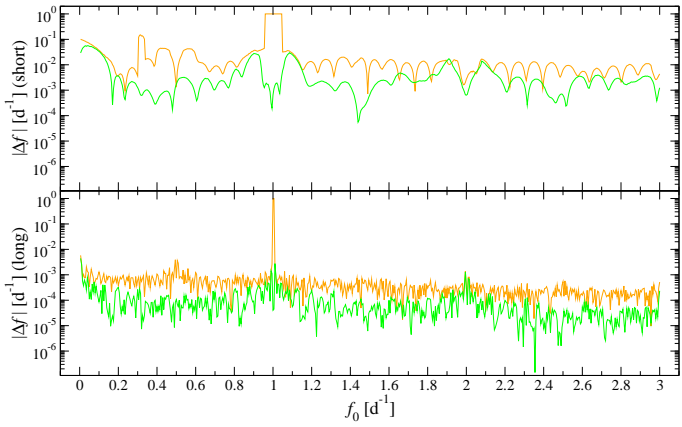}
\caption{Frequency accuracy (rms scatter of resulting frequencies about initial frequency for a single sinusoidal signal with uniformly distributed phase) vs.~signal frequency of DFT ({\em orange}) and Lomb-Scargle Periodogram ({\em green}). The simulated time series represent pure signal without noise at uniformly distributed phase angles. The {\em top} panel represents the frequency accuracy for seven consecutive nights of single-site photometry of 44\,Tau (Str\o mgren $y$, 619 data points). These data are a subset of the $81$ days long time series presented in the {\em bottom} panel (34 nights, 2\,981 data points). The plots illustrate the improvement of the overall frequency accuracy of both methods for the longer dataset. The corresponding deviations obtained using the spectral significance are practically zero, similar to the top panel in Fig.\,\ref{FIGfacc}, and thus not presented here. Only those attempts where the distance between resulting frequency and input frequency does not exceed $0.5$\,d$^{-1}$ were taken into account, the rest was considered alias. The results are based on a numerical simulation investigating $1\, 000$ datasets for every frequency.}\label{FIGsamplesize}
\end{figure}

Fig.\,\ref{FIGsamplesize} shows the frequency accuracies of the compared methods for the short dataset in the {\em top} panel. The display is exactly according to Fig.\,\ref{FIGfacc}. The {\em bottom} panel contains the identical analysis of frequency accuracies for the long dataset and indicates an improved overall precision of the DFT and Lomb-Scargle Periodogram, if the number of data increases.

\begin{figure}\includegraphics[width=256pt]{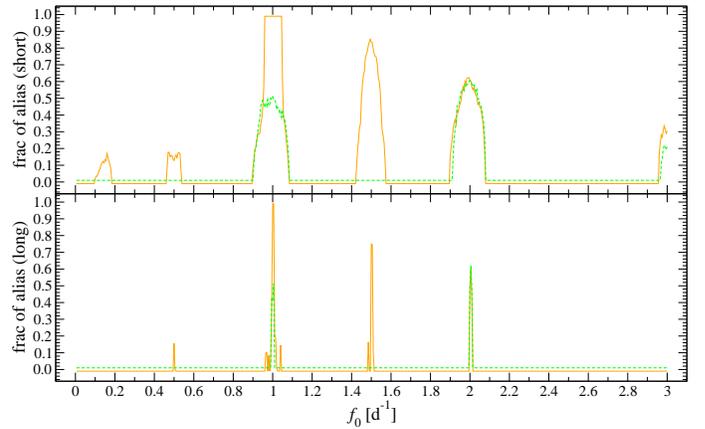}
\caption{Same as Fig.\,\ref{FIGsamplesize}, but displaying the relative number of aliases vs.~the signal frequency of DFT ({\em solid orange}) and Lomb-Scargle Periodogram ({\em dashed green}). In this context, a result is considered alias, if the absolute difference between resulting frequency and signal frequency exceeds $0.5$\,d$^{-1}$. The spectral significance does not produce a single alias maximum in the simulations, similar to the top panel in Fig.\,\ref{FIGalias}. The corresponding graphs are thus not presented here. The DFT graph is offset vertically by $-0.01$ and the Lomb-Scargle graph by $+0.01$ to provide better visibility at and close to zero.}\label{FIGssalias}
\end{figure}

However, 1\,d$^{-1}$ aliasing persists also for the long time series (Fig.\,\ref{FIGssalias}), which is compatible with the persisting alias peaks in the spectral window at integer multiples of 1\,d$^{-1}$. Since the zero-mean correction is represented by the subtraction of a constant from all observables in the entire time series (referring to eqs.\,\ref{EQ a_ZM}, \ref{EQ b_ZM}), it is not surprising that the techniques which ignore the statistical effects of zero-mean correction retain these spectral window-related weaknesses also in case of a large number of data points.

The result of spectral significance calculations is not provided in these graphs: the deviation of frequencies is practically equal to zero independently of frequency -- as for the IC\,4996\,\#\,89 data (top panel in Fig.\,\ref{FIGfacc}).

\section{{\sc SigSpec}: Practical application}\label{S Practical application}

The formal concept of spectral significance does not only provide reliable information on the sampling characteristics in the time domain, but also allows to include the computations into a prewhitening sequence. {\sc SigSpec} realizes the entire procedure as a high-performance algorithm. Implementations for various operating systems and CPUs for free download at {\tt http://www.astro.univie.ac.at/SigSpec}. The ANSI-C source code is available on request ({\tt reegen@astro.univie.ac.at}).

The {\sc SigSpec} technique was applied to numerous datasets and has proven its value also in other scientific problems. A practical example for a situation where {\sc SigSpec} performs superior to classical DFT-based signal-to-noise ratio estimation is given in Table\,\ref{TAB44Tau}. Columns 1 and 2 represent 13 identified eigenfrequencies plus $16$ excited linear combinations of the $\delta$\,Sct star 44\,Tau (Antoci et al.~2006). The corresponding Str\o mgren $v$ and $y$ amplitudes as obtained from a multisite campaign in 2001/02 are displayed in columns 3 and 4. The underlying light curves consist of 3\,890 points in $v$ and 3\,582 points in $y$. For the analysis of the campaign data, the signal-to-noise ratio threshold for peak acceptance was chosen to be $3.5$.

\begin{table}
\caption{Frequencies detected in multisite data of 44\,Tau, according to Antoci et al.~(2006). $A_v$ and $A_y$ are the published amplitudes (mmag) in Str\o mgren $v$ and $y$ for the 2001/02 multisite data (3\,890 points in $v$, 3\,582 points in $y$). The corresponding columns indicated by `S/N' contain a cross-identification for the data obtained by the Vienna University Automatic Photoelectric Telescope (Strassmeier et al.~1997; Granzer et al.~2001) in the course of this campaign (3\,280 points in $v$, 2\,981 points in $y$). An amplitude signal-to-noise ratio $> 4$ was used. The columns indicated by `sig' represent the cross-identification of the Vienna APT data with the complete campaign dataset using the result of the {\sc SigSpec} analysis, based on a spectral significance limit of $5.46$. Amplitudes in {\it italic} print refer to 1\,d$^{-1}$ alias peaks. Since the peaks for the APT subset were assigned to corresponding peaks in the total dataset, individual frequencies are not displayed.}\label{TAB44Tau}
\begin{scriptsize}
\begin{tabular}{lrrrrrrr}
             & $f$ $\left[\mathrm{d}^{-1}\right]$                            & \multicolumn{2}{c}{total} & \multicolumn{2}{c}{APT, S/N} & \multicolumn{2}{c}{APT, sig} \\
\cline{3-4} \cline{5-6} \cline{7-8} \\
id&&\multicolumn{1}{c}{$A_v$}&\multicolumn{1}{c}{$A_y$}&\multicolumn{1}{c}{$A_v$}&\multicolumn{1}{c}{$A_y$}&\multicolumn{1}{c}{$A_v$}&\multicolumn{1}{c}{$A_y$}\\
$f_1$     & 6.8980&39.51&27.27&39.41&27.20&39.95&27.14\\
$f_2$     & 7.0060&17.20&12.15&18.90&13.21&19.95&13.61\\
$f_3$     & 9.1175&21.02&14.57&16.96&11.62&16.34&11.24\\
$f_4$     &11.5196&16.56&11.79&18.28&12.88&17.56&12.90\\
$f_5$     & 8.9606&13.73& 9.32&13.86& 9.80&14.63& 9.75\\
$f_6$     & 9.5613&18.62&12.92&10.84& 7.38& 9.94& 6.68\\
$f_7$     & 7.3034& 5.46& 3.76& 7.23& 4.83& 7.37& 5.06\\
$f_8$     & 6.7953& 4.83& 3.27& 3.77& 2.79& 4.32& 3.03\\
$f_9$     & 9.5801& 3.64& 2.25& 1.97& 1.28& 2.23& 1.53\\
$f_{10}$  & 6.3390& 2.08& 1.62& 2.34& 1.78& 2.25& 1.86\\
$f_{11}$  & 8.6394& 1.71& 1.45& 2.00& 1.56& 1.72& 1.21\\
$f_{12}$  &11.2946& 1.35& 0.92& 1.02&  ---&{\it 0.72}&  ---\\
$f_{13}$  &12.6967& 0.24& 0.52&  ---&  ---&{\it 0.59}&  ---\\
\\
\hline
\\
$2f_1$    &13.7962& 1.58& 1.18& 1.52& 1.38& 1.65& 1.32\\
$f_1+f_2$ &13.9040& 1.45& 1.02& 1.35&{\it 1.02}& 1.24 &{\it 0.98}\\
$2f_2$    &14.0120& 0.40& 0.50&  ---&  ---&{\it 0.80} &  ---\\
$f_1+f_5$ &15.8586& 1.16& 1.02& 1.22&  ---& 1.22&{\it 0.80}\\
$f_1+f_3$ &16.0155& 1.63& 1.29&{\it 0.93}&  --- &{\it 0.69}&{\it 0.64}\\
$f_2+f_3$ &16.1235& 0.79& 0.85&  ---&  ---&{\it 0.43}& 0.41\\
$f_1+f_6$ &16.4593& 1.25& 0.66&  ---&  ---&{\it 0.36}&  ---\\
$f_2+f_6$ &16.5673& 0.65& 0.32&  ---&  ---&  ---& 0.46\\
$f_4+f_8$ &18.3149& 1.11& 0.48&  ---&  ---& 0.56& 0.55\\
$f_1+f_4$ &18.4177& 1.69& 1.10&{\it 1.67}&{\it 1.27}&{\it 1.95}&{\it 1.45}\\
$f_2+f_4$ &18.5256& 1.14& 0.96& 0.96&  ---& 0.98& 0.59\\
$f_4+f_5$ &20.4802& 1.22& 0.94& 1.24& 1.07& 1.24& 1.09\\
$f_3+f_4$ &20.6371& 0.91& 0.81&  ---&  ---& 0.40&  ---\\
$f_4+f_6$ &21.0809& 0.77& 0.46&  ---&  ---&{\it 0.58}& 0.46\\
$2f_4$    &23.0392& 1.12& 0.78& 1.17&  ---& 1.11& 0.78\\
$2f_4+f_1$&29.9373& 0.57& 0.35&  ---&  ---& 0.70& 0.50\\
\end{tabular}
\end{scriptsize}
\end{table}

The major part of the data (3\,280 points in $v$, 2\,981 points in $y$) was acquired by the Vienna University Automatic Photoelectric Telescope (APT; Strassmeier et al.~1997; Granzer et al.~2001). The data represent 34 nights covering a time interval of $81$\,d. As a comparative test of the practical performance of {\sc SigSpec}, the APT data alone were analysed using a DFT-based prewhitening sequence relying on a signal-to-noise ratio limit of $4$. Except for the higher threshold, this is exactly the same technique as applied to the campaign data. In any case, the frequencies found in the APT subset were cross-identified with the frequencies published for the full dataset (Antoci et al.~2006). The result for the APT data is displayed in columns 5 and 6. In $v$, $20$ of the $29$ signal components are found, in $y$ only $15$. With each filter, 1\,d$^{-1}$ aliasing is found for $2$ frequencies, which are indicated by {\it italc} print.

Columns 7 and 8 refer to the corresponding {\sc SigSpec} analysis using a spectral significance limit of $5.46$ and reproducing all frequencies except for one in $v$ and $5$ in $y$. The number of aliases is $8$ in $v$ and $4$ in $y$, but the majority of aliases in the {\sc SigSpec} result occurs for frequencies not resolved by the alternative technique at all.

For additional practical examples, the reader is referred to Reegen (2005).

\section{Further Topics}\label{S Further topics}

\subsection{Spectral significance for statistically weighted time series}\label{SUB Statistical weights}

Astronomical measurements are generally influenced by instrumental and environmental conditions changing with time. Thus different accuracies for different data points are involved, which are frequently desired to be taken into account by applying weights to the observables. If the accuracy is poor, the corresponding weight is low, and high-accuracy data points are assigned a high weight, respectively. Furthermore, multisite campaigns employ different telescopes with different instrumental parameters, which may also require an appropriate weighting.

This section refers to a set of statistical weights, $\gamma _k$, $k = 0,\, ...\, K-1$, normalized according to
\begin{equation}\label{EQ weights}
\sum _{k=0}^{K-1}\gamma _k =: K\: .
\end{equation}

Weighting may be considered as counting an item $\xi _k$ of, e.\,g., an arithmetic mean repeatedly instead of only once. If the number of counts for an arbitrary $\xi _k$ is $n_k$, then the arithmetic mean over all $\xi _k$, $k=0,1,...,K-1$ is expressed as
\begin{equation}
\left<\xi _k\right> = \frac{\sum _{k=0}^{K}n_k\xi _k}{\sum _{k=0}^{K}n_k}\: .
\end{equation}
In this case, eq.\,\ref{EQ weights} demands the normalized weights to be introduced according to
\begin{equation}
\gamma _k := \frac{K\xi _k}{\sum _{k=0}^{K}n_k}\: .
\end{equation}
These arguments may consistently be followed to obtain the subsequent relations (eqs.\,\ref{EQ weighted mean} to \ref{EQ phases of extreme variance weights}).

The implementation of statistical weights has to be performed for all elements of the spectral significance relation (eq.\,\ref{EQ significance full}, or \ref{EQ significance full cartesian}, respectively).
\begin{enumerate}
\item The weighted mean value to be subtracted from every data point is calculated according to
\begin{equation}\label{EQ weighted mean}
\left< x_k\right> = \frac{1}{K}\sum _{k=0}^{K-1}\gamma _k x_k\: .
\end{equation}
Further considerations assume $x_k$ to be already zero-mean corrected.
\item Also the weighted mean variance of the data,
\begin{equation}
\left< x_k^2\right> = \frac{1}{K}\sum _{k=0}^{K-1}\gamma _k x_k^2\: ,
\end{equation}
can be used as an estimator for the variance of the corresponding population.
\item The DFT is determined by the weighted Fourier Coefficients
\begin{eqnarray}
a_{\mathrm{ZM}}\left(\omega _n\right) & := & \frac{1}{K}\sum_{k=0}^{K-1}\gamma _k x_k\cos\omega _nt_k\: ,\\
b_{\mathrm{ZM}}\left(\omega _n\right) & := & \frac{1}{K}\sum_{k=0}^{K-1}\gamma _k x_k\sin\omega _nt_k\: .
\end{eqnarray}
\item
The parameters characterizing the sampling properties may be generalized according to
\begin{eqnarray}
\nonumber &&\alpha _0^2\left(\omega ,\theta _0\right) = \frac{2}{K^2}\\
\label{EQ normalized semi-major axis weights}&&\left\lbrace K\sum_{k=0}^{K-1}\gamma _k\cos ^2\left(\omega t_k - \theta _0\right) - \left[\sum_{k=0}^{K-1}\gamma _k\cos\left(\omega t_l - \theta _0\right)\right] ^2\right\rbrace\: ,\\
\nonumber &&\beta _0^2\left(\omega ,\theta _0\right) = \frac{2}{K^2}\\
\label{EQ normalized semi-minor axis weights}&&\left\lbrace K\sum_{k=0}^{K-1}\gamma _k\sin ^2\left(\omega t_k - \theta _0\right) - \left[\sum_{k=0}^{K-1}\gamma _k\sin\left(\omega t_l - \theta _0\right)\right] ^2\right\rbrace\: , \\
\nonumber &&\tan 2\theta _0\left(\omega\right) =\\
\nonumber &&\left[ K\sum_{k=0}^{K-1}\gamma _k\sin 2\omega t_k - 2\sum_{k=0}^{K-1}\gamma _k\cos\omega t_k\sum_{k=0}^{K-1}\gamma _k\sin\omega t_k\right]\: \\
\nonumber &&\cdot\left[ K\sum_{k=0}^{K-1}\gamma _k\cos 2\omega t_k - \left(\sum_{k=0}^{K-1}\gamma _k\cos\omega t_k\right) ^2\right.\\
\label{EQ phases of extreme variance weights}&& + \left.\left(\sum_{k=0}^{K-1}\gamma _k\sin\omega t_k\right) ^2\right] ^{-1}\: .
\end{eqnarray}
\end{enumerate}
These parameters provide eq.\,\ref{EQ significance full cartesian} to return the weighted significance spectrum of the dataset.

In the case of $\gamma _k =: 1$ $\forall k$, these relations consistently reduce to eqs.\,\ref{EQ phases of extreme variance}, \ref{EQ normalized semi-major axis}, \ref{EQ normalized semi-minor axis} for unweighted analysis, respectively. Given the above re-definitions, eq.\,\ref{EQ significance full} and the corresponding considerations -- conversion between signal-to-noise ratio and signficance (\ref{SUBSUB Connection between significance and signal-to-noise ratio}) and Sock Diagram (\ref{SUBSUB The Sock Diagram}) -- readily apply to weighted time series as well.

\subsection{Spectral significance for zero-mean corrected subsets}\label{SUB Spectral significance for zero-mean corrected subsets}

In some cases one may demand to correct the mean magnitude for individual subsets instead of the entire time series, i.\,e. if the data were obtained by different instruments, or if the subtraction of nightly averages is performed. In this case the influence on the statistical description of the noise to which an amplitude level is compared has to reflect the subdivision correspondingly. Given a subset index, $s = 1,2,...,S$, where $S$ denotes the total number of subsets, eqs.\,\ref{EQ a_ZM}, \ref{EQ b_ZM} have to be re-written according to
\begin{eqnarray}
\nonumber &&a_{\mathrm{ZM}}\left(\omega\right) =\\
&&\frac{1}{K}\left[\sum_{s=1}^{S}\left(\sum_{k=0}^{K_s-1}x_{sk}\cos\omega t_{sk} - \frac{1}{K_s}\sum_{k=0}^{K_s-1}x_{sk}\sum_{l=0}^{K_s-1}\cos\omega t_{sl}\right)\right] \: ,\\
\nonumber &&b_{\mathrm{ZM}}\left(\omega\right) =\\ &&\frac{1}{K}\left[\sum_{s=1}^{S}\left(\sum_{k=0}^{K_s-1}x_{sk}\sin\omega t_{sk}-  \frac{1}{K_s}\sum_{k=0}^{K_s-1}x_{sk}\sum_{l=0}^{K_s-1}\sin\omega t_{sl}\right)\right] \: .
\end{eqnarray}
In this context, each registration time and observable is assigned two indices, writing $x_{sk}$. The first index refers to the subset, and the second index is the identifier of the data point within the subset. The number of data points within the subset $s$ is denoted by $K_s$, and finally $K := \sum_{s=1}^{S}K_s$ for the total number of data points in the entire time series.

Further calculations follow the procedure described in Section\,\ref{S Amplitude PDF for non-equidistant sampling} exactly and yield
\[
\tan 2\theta _0\left(\omega\right) =
\left[\sum_{s=1}^{S}K_s\sum_{k=0}^{K_s-1}\sin 2\omega t_{sk}-\sum _{k=0}^{K_s-1}\cos\omega t_{sk}\sum_{k=0}^{K_s-1}\sin\omega t_{sk}\right]\:
\]
\begin{equation}
\cdot\left[\sum_{s=1}^{S}K_s\sum_{k=0}^{K_s-1}\cos 2\omega t_{sk}-\left(\sum_{k=0}^{K_s-1}\cos\omega t_{sk}\right) ^2+\left(\sum_{k=0}^{K_s-1}\sin\omega t_{sk}\right) ^2\right] ^{-1}\!\!\!
\end{equation}
for the orientation of the rms error ellipse, in analogy to eq.\,\ref{EQ phases of extreme variance}. Correspondingly, the normalized axes of the ellipse (eqs.\,\ref{EQ normalized semi-major axis}, \ref{EQ normalized semi-minor axis}) evaluate to
\begin{eqnarray}
\nonumber &&\alpha _0\left(\omega ,\theta _0\right) =\left(\frac{2}{K}\sum_{s=1}^{S}\left\lbrace\sum_{k=0}^{K_s-1}\cos ^2\left(\omega t_{sk} - \theta _0\right)\right.\right.\\
&&- \left.\left.\frac{1}{K_s}\left[\sum_{l=0}^{K_s-1}\cos\left(\omega t_{sl} - \theta _0\right)\right] ^2\right\rbrace\right) ^{\frac{1}{2}}\: ,\\
\nonumber &&\beta _0\left(\omega ,\theta _0\right) =\left(\frac{2}{K}\sum_{s=1}^{S}\left\lbrace\sum_{k=0}^{K_s-1}\sin ^2\left(\omega t_{sk} - \theta _0\right)\right.\right.\\
&&- \left.\left.\frac{1}{K_s}\left[\sum_{l=0}^{K_s-1}\sin\left(\omega t_{sl} - \theta _0\right)\right] ^2\right\rbrace\right) ^{\frac{1}{2}}\: .
\end{eqnarray}

\subsection{Colored noise}\label{SUB Colored noise}

{\sc SigSpec} does not take into account colored noise. Since there are both instrumental effects (e.\,g.~CCD readout, stability of the spacecraft position) and stellar variations (e.\,g.~low-amplitude modes, granulation noise) to be considered and neither of these two sources may be determined unambiguously, it is presently impossible to deduce a reliable amplitude noise profile for measurements in many cases.

The heuristic approach to generate a noise spectrum by means of (weighted) moving averages suffers from the presence of unresolved peaks, which increases the risk to miss intrinsic signal. Hence it is advisable to use {\sc SigSpec} for the detection of significant sinusoidal signals and to take effects due to colored noise (however the latter may have been determined) into account by choosing different spectral significance thresholds for different frequency regions.

However, there is evident demand for further investigation in the presence of colored noise. The statistical description of colored noise and its implementation into the evaluation of the spectral significance is the subject of present and ongoing investigation and will be discussed in a dedicated publication (paper III).

\section{Conclusions}\label{S Conclusions}

{\sc SigSpec} exceeds the diagnostic capabilities of Discrete Fourier Transform, Lomb-Scargle Periodogram, Phase Dispersion Minimization, and least-squares fits in various respects.

\begin{enumerate}
\item {\sc SigSpec} does not compare one peak amplitude to another. Thus it avoids implications to the statistical independence of peaks in the amplitude spectrum at a single prewhitening stage and the ambiguity in defining a Nyquist Frequency, which all compared methods suffer from. The implicit comparison of peaks by selecting the dominant one for prewhitening may be performed free of the frequency and phase dependencies of Fourier amplitudes, if the spectral significance is used instead (\ref{SUB Comparison of analytical and numerical solutions}). One remaining issue in this context is the effect of multiple hypothesis testing on step-by-step identification and prewhitening. The propagation of uncertainties from one prewhitening step to another will be a topic in Paper II examining the DFT on multi-periodic signals.
\item {\sc SigSpec} is based on an analytically straight determination of the amplitude Probability Density Function (\ref{SUB Frequency- and phase-dependent PDF}). Provided the noise is white (but not necessarily Gaussian), {\sc SigSpec} returns {\em exact} results instead of signal-to-noise ratio estimates (\ref{SUB Comparison of analytical and numerical solutions}). A more realistic approach is the introduction of colored noise in the sense of serially correlated measurements. Paper III is planned to discuss the appropriate incorporation of the serial correlation into the spectral significance computation.
\item {\sc SigSpec} is the first technique in astronomical time series analysis to use both frequency and phase angle in the computation of the False-Alarm Probability (Section\,\ref{S Amplitude PDF for non-equidistant sampling}), appropriately encountering the fact that the amplitude noise level is systematically different for different frequencies and phases in Fourier Space (\ref{SUB Accuracy of peak frequencies}). The statistics employed by {\sc SigSpec} takes into account {\em all available} information provided by the Discrete Fourier Transform.
\item The performance of {\sc SigSpec} for a single signal plus noise is discussed in this paper  (\ref{SUB Accuracy of peak frequencies}, \ref{SUB Aliasing}). For strong signal components, the peak frequencies returned by {\sc SigSpec} are considerably more accurate than the solutions of all compared methods. For weak signal components, {\sc SigSpec} still provides a slightly higher accuracy. Furthermore, there is indication that the {\sc SigSpec} frequencies are at least as accurate as the least-squares fits. The systematic examination of multiperiodic signals and a comparison to non-linear least-squares fits will be provided in paper II.
\item Signal-to-noise ratio estimation provides a valid significance estimate, but becomes very poor for low frequencies and in case of periodicities in the time-domain sampling (\ref{SUB Sample size effects}).
\item Signal-to-noise ratio-based period detection is a matter of experience and -- to some extent -- personal taste. {\sc SigSpec} provides a solution that is completely free of subjective influence. Interpretation is no longer a part of the analysis, and the results remain exactly the same, if two different persons apply the calculations to the same data.
\end{enumerate}
Tests on synthetic and real data confirm that the {\sc SigSpec} analysis is superior to signal-to-noise ratio estimates (Section\,\ref{S Numerical tests}).

The Sock Diagram (\ref{SUBSUB The Sock Diagram}), as a generalized analogy to the spectral window, provides increased information on the properties of the time-domain sampling in Fourier Space, since incorporating phase-resolution as well. Periodicities in the sampling of a time series are recovered by the spectral window and provide a measure of the susceptibility of DFT to aliasing. In addition, the Sock Diagram indicates frequency and phase regions with high systematic errors of DFT amplitudes. Both error sources may cause the classical signal-to-noise ratio estimation to fail.

The incorporation of statistical weights into the spectral significance calculation is provided  (\ref{SUB Statistical weights}), as well as the modification of the spectral significance computation, if a time series is assumed to consist of several zero-mean corrected subsets (\ref{SUB Spectral significance for zero-mean corrected subsets}). Both features also implemented in the present software version.

The {\sc SigSpec} software has proved its validity and excellent performance in numerous practical applications, mainly in connection with the data from the MOST satellite (Walker et al.~2005; Guenther et al.~2005), but also for the photometric search for new pre-main sequence Delta Scuti stars in young open clusters (Zwintz et al.~2004; Zwintz \& Weiss~2006). In addition, {\sc SigSpec} was successively employed for the photometric and spectroscopic determination of the rotation periods of selected roAp stars (Sachkov et al.~2004; Ryabchikova et al.~2005).

However, although the statistically unbiased approach used by {\sc SigSpec} improves the reliability of results considerably, the software does not provide an invitation to blind belief. It successfully separates the detection of {\em formally} significant sinusoidal signal components in a time series from their interpretation in a physical context, nothing more. The practical application of the program still requires a careful examination of results, namely the decision whether a formally significant frequency is related to the observed physical process or an (e.\,g. instrumental) artifact.

\subsection*{Acknowledgements}

PR received financial support from the Fonds zur F\"orderung der wis\-sen\-schaft\-li\-chen Forschung (FWF, projects P14546-PHY, P14984-PHY) Furthermore, it is a pleasure to thank T.~Appourchaux (IAS, Orsay), A.~Baglin (Obs.~de Paris, Meudon), T.~Boehm (Obs.~M.-P., Tou\-lou\-se), M.~Breger, R.~Dvorak, M.\,G.~Firneis, D.~Frast (Univ.~of Vienna), R.~Garrido (Inst.~Astrof.~Andalucia, Granada), M.~Gruberbauer (Univ.~of Vienna), D.\,B.~Guenther (St.~Mary's Univ., Halifax), M.~Hareter, D.~Huber, T.~Kallinger (Univ.~of Vienna), R.~Kusch\-nig (UBC, Vancouver), S.~Mar\-chen\-ko (Western Kentucky Univ., Bowling Green, KY), M.~Masser (Univ.~of Vienna), J.\,M. Matthews (UBC, Vancouver), E.~Michel (Obs. de Paris, Meudon), A.\,F.\,J.~Moffat (Univ.~de Montreal), E.~Paunzen, D.~Punz (Univ.~of Vienna), V.~Ripepi (INAF, Naples), S.\,M. Rucinski (D.~Dunlap Obs., Toronto), T.\,A.~Ryabchikova (Inst.~Astr. RAS, Moscow), D.~Sasselov (Harvard-Smithsonian Center, Cambridge, MA), S.~Schraml (Univ.~of Technology, Vienna), G.\,A.~Wade (Royal Military College, Kingston), G.\,A.\,H.~Walker (UBC, Vancouver), W.\,W. Weiss, and K.~Zwintz (Univ.~of Vienna) for valuable discussion and support with extensive software tests. Moreover, M.\,G.~Firneis, D.\,B.~Guenther, J.\,M. Matthews, G.\,A.\,H.~Walker, and W.\,W. Weiss were a great help in clarifying the exposition of this paper.

Finally, I address my very special thanks to the referee of this paper, J.\,D.~Scargle, for his contribution that substantially improved the overall quality of the publication.

\appendix\onecolumn

\section{Orientation of the rms Error Ellipse of Fourier Coefficients}\label{APPENDIX Orientation of the rms Error Ellipse of Fourier Coefficients}

Rotating the coordinate system in Fourier Space by a phase angle $\theta _0$ transforms $a_\mathrm{ZM}$, $b_\mathrm{ZM}$ from eqs.\,\ref{EQ zero-mean a}, \ref{EQ zero-mean b} into
\begin{eqnarray}
\alpha\left(\omega ,\theta _0\right) & = & \frac{1}{K}\sum_{k=0}^{K-1}x_k\left[\cos\left(\omega t_k-\theta _0\right) - \frac{1}{K}\sum_{l=0}^{K-1}\cos\left(\omega t_l-\theta _0\right)\right]\: ,\\
\beta\left(\omega ,\theta _0\right) & = & \frac{1}{K}\sum_{k=0}^{K-1}x_k\left[\sin\left(\omega t_k-\theta _0\right) - \frac{1}{K}\sum_{l=0}^{K-1}\sin\left(\omega t_l-\theta _0\right)\right]\: .
\end{eqnarray}
In terms of the population variance $\left< x^2\right>$ of the time-domain data $x_k$, the population variances of $\alpha$ and $\beta$ are
\begin{eqnarray}
\label{EQ CosCoef extreme variance}\left<\alpha ^2\right>\left(\omega ,\theta _0\right) & = & \frac{\left< x^2\right>}{K^3}\left\lbrace K\sum_{k=0}^{K-1}\cos ^2\left(\omega t_k - \theta _0\right) - \left[\sum_{l=0}^{K-1}\cos\left(\omega t_l - \theta _0\right)\right] ^2\right\rbrace\: ,\\
\label{EQ SinCoef extreme variance}\left<\beta ^2\right>\left(\omega ,\theta _0\right) & = & \frac{\left< x^2\right>}{K^3}\left\lbrace K\sum_{k=0}^{K-1}\sin ^2\left(\omega t_k - \theta _0\right) - \left[\sum_{l=0}^{K-1}\sin\left(\omega t_l - \theta _0\right)\right] ^2\right\rbrace\: .
\end{eqnarray}
The covariance
\begin{equation}
\left<\alpha\beta\right> = \frac{1}{K}\sum_{k=0}^{K-1}x_k^2\left[\cos\left(\omega t_k-\theta _0\right) - \frac{1}{K}\sum_{l=0}^{K-1}\cos\left(\omega t_l-\theta _0\right)\right]\left[\sin\left(\omega t_k-\theta _0\right) - \frac{1}{K}\sum_{l=0}^{K-1}\sin\left(\omega t_l-\theta _0\right)\right]
\end{equation}
vanishes for a random population, if
\begin{equation}\label{EQ cov 0}
\sum_{k=0}^{K-1}\left[\cos\left(\omega t_k-\theta _0\right) - \frac{1}{K}\sum_{l=0}^{K-1}\cos\left(\omega t_l-\theta _0\right)\right]\left[\sin\left(\omega t_k-\theta _0\right) - \frac{1}{K}\sum_{l=0}^{K-1}\sin\left(\omega t_l-\theta _0\right)\right] = 0\: .
\end{equation}
The constant $\theta _0$ may be separated from the sums, which yields
\[
\left(\cos ^2\theta _0 - \sin ^2\theta _0\right)\left(K\sum_{k=0}^{K}\cos\omega t_k\sin\omega t_k - \sum_{k=0}^{K}\cos\omega t_k\sum_{k=0}^{K}\sin\omega t_k\right) =
\]
\begin{equation}\label{EQ Appendix A 1}
\cos\theta _0\sin\theta _0\left[K\sum_{k=0}^{K}\left(\cos ^2\omega t_k-\sin ^2\omega t_k\right)-\left(\sum_{k=0}^{K}a\cos\omega t_k\right) ^2+\left(\sum_{k=0}^{K}\sin\omega t_k\right) ^2\right]\: ,
\end{equation}
where it is not necessary to distinguish between indices $k$ and $l$ any more. Eq.\,\ref{EQ Appendix A 1} may be expressed in terms of $2\,\theta _0$ and $2\,\omega t_k$, respectively, according to
\[
\cos 2\,\theta _0\left[K\sum_{k=0}^{K}\sin 2\omega t_k-2\sum_{k=0}^{K}\cos\omega t_k\sum_{k=0}^{K}\sin\omega t_k\right] =
\]
\begin{equation}
\sin 2\,\theta _0\left[K\sum_{k=0}^{K}\cos 2\omega t_k-\left(\sum_{k=0}^{K}\cos\omega t_k\right) ^2+\left(\sum_{k=0}^{K}\sin\omega t_k\right) ^2\right]\: .
\end{equation}
This immediately leads to eq.\,\ref{EQ phases of extreme variance}.

\section{PDF Transformation}\label{APPENDIX PDF Transformation}

\subsection{General PDF transformation}\label{SUB General PDF transformation}

According to, e.\,g., Stuart \& Ord (1994), pp.\,20ff, the PDF conversion between a random variable $x$ and a random variable $y$ connected via $y := y\left( x\right)$ is given by
\begin{equation}\label{EQ PDF transformation}
\phi_x\left( x\right) = \phi_y\left( y\right)\,\left|\frac{dy}{dx}\right|\: .
\end{equation}

For multivariate distributions, the generalization of differentials yields
\begin{equation}
\phi_x\left( x\right) = \phi_y\left( y\right)\,\left|\frac{d\vec{y}}{d\vec{x}}\right|\: ,
\end{equation}
with the Jacobian
\begin{equation}\label{EQ multivariate PDF transformation}
\left|\frac{d\vec{y}}{d\vec{x}}\right| := \left|
\begin{array}{cccc}
\frac{\partial y_1}{\partial x_1} & \frac{\partial y_2}{\partial x_1} & \frac{\partial y_3}{\partial x_1} & \dots \\ 
\frac{\partial y_1}{\partial x_2} & \frac{\partial y_2}{\partial x_2} & \frac{\partial y_3}{\partial x_2} & \dots \\ 
\frac{\partial y_1}{\partial x_3} & \frac{\partial y_2}{\partial x_3} & \frac{\partial y_3}{\partial x_3} & \dots \\ 
\vdots & \vdots & \vdots & \ddots
\end{array}
\right|\: .
\end{equation}

\subsection{Example: transformations of PDF and expected value for the exponential distribution}\label{SUB Example: transformations of PDF and expected value for the exponential distribution}

If $y$ is an exponentially distributed random variable with the PDF
\begin{equation}\label{EQ exponential PDF}
\phi_y\left( y\right) =: \kappa\,\mathrm{e}^{-\kappa y}\: ,
\end{equation}
eq.\,\ref{EQ PDF transformation} for $x:=\sqrt{y}$ evaluates to
\begin{equation}\label{EQ transformed exponential PDF}
\phi_x\left( x\right) = 2\kappa x\,\mathrm{e}^{-\kappa x^2}\: ,
\end{equation}
where both $x$, $y$ are considered positive semi-definite.

The expected value of $y$ is
\begin{equation}\label{EQ exponential expected value}
\left< y\right> =: \int _{0}^{\infty}dy\, y\phi_y\left( y\right) = \frac{1}{\kappa}\: ,
\end{equation}
as obtained from eq.\,\ref{EQ exponential PDF}. Correspondingly, eq.\,\ref{EQ transformed exponential PDF} yields
\begin{equation}\label{EQ transformed exponential expected value}
\left< x\right> =: \int _{0}^{\infty}dx\, x\phi_x\left( x\right) = \sqrt{\frac{\pi}{2\,\kappa}}\: ,
\end{equation}
so that
\begin{equation}\label{EQ transformation exponential expected value}
\left< y\right> = \frac{4}{\pi}\,\left< x\right> ^2
\end{equation}
describes the transformation of expected values corresponding to $y = x^2$.

For equidistantly sampled noise $x_k$ in the time domain, the power spectrum associated with the Fourier Series is exponentially distributed (Schuster 1898; Scargle 1982) according to
\begin{equation}\label{EQ PDF Schuster}
\phi\left( A^2\right) = \frac{K}{2\left< x^2\right>}\,\mathrm{e}^{-\frac{KA^2}{2\left< x^2\right>}}\: ,
\end{equation}
where $\left< x^2\right>$ denotes the population variance of the noise. The previous considerations apply through the substitution $\kappa =: \frac{K}{2\left< x^2\right>}$, normalized to single-sided power spectral density (which introduces the factor $2$). According to eq.\,\ref{EQ exponential expected value}, the noise level in the spectrum of squared amplitudes evaluates to
\begin{equation}
\left< A^2\right> = \frac{2\left< x^2\right>}{K}\: .
\end{equation}
For the expected value of the amplitude (i.\,e.~the amplitude noise level), eq.\,\ref{EQ transformed exponential expected value} yields
\begin{equation}\label{EQ amplitude noise}
\left< A\right> = \sqrt{\frac{\pi}{K}\,\left< x^2\right>}\: .
\end{equation}

Eq.\,\ref{EQ transformation exponential expected value} leads to the conversion between the expected values of amplitude, $\left< A\right>$ and squared amplitude, $\left< A^2\right>$, according to
\begin{equation}\label{EQ amp pow noise}
\left< A^2\right> = \frac{4}{\pi}\,\left< A\right> ^2.
\end{equation}
This is in agreement with the observational results by Horne \& Baliunas (1986).

An amplitude signal-to-noise ratio of $4$ -- introduced as an approximate significance criterion by Breger et al.~(1993) -- corresponds to a power signal-to-noise ratio of $4\,\pi\approx 12.57$.

\end{document}